\documentclass[twocolumn]{aa} 

\usepackage{txfonts}
\usepackage[utf8]{inputenc}
\usepackage{dirtytalk}
\usepackage{amsmath}
\usepackage{amsfonts}
\usepackage{amssymb}
\usepackage{soul}
\usepackage{graphicx}
\usepackage{subcaption}
\usepackage{bm}
\usepackage{latexsym}
\usepackage{epsfig}
\usepackage[normalem]{ulem}
\usepackage[colorlinks=true, citecolor=blue]{hyperref}
\usepackage[dvipsnames]{xcolor}
\usepackage{todonotes}
\hypersetup{colorlinks=true,allcolors=blue}
\usepackage{orcidlink}
\def\aap{A\&A}

\def\apjs{ApJS}
\def\apjl{ApJL}

\newcommand{\Msun}{\mathrm{M}_\odot}

\newcommand{\kmsMpc}{km~s$^{-1}$~Mpc$^{-1}$}
\newcommand{\gpcyr}{Gpc$^{-3}$~yr$^{-1}$}

\newcommand{\gwu}{\texttt{GWSim}}
\newcommand{\mcat}{\texttt{MICECATv2}}
\newcommand{\mcatv}{\texttt{MICECATv2}}
\newcommand{\lcdm}{$\Lambda$CDM}
\newcommand{\xeff}{\chi_\text{eff}}
\newcommand{\dd}{\text{d}}
\newcommand{\der}[2]{\frac{\dd{#1}}{\dd{#2}}}
\newcommand{\dder}[3]{\frac{\dd{#1}}{\dd{#2}\,\dd{#3}}}

\newcommand{\beq}{\begin{equation}}
\newcommand{\eeq}{\end{equation}}

\newcommand{\eq}{Eq.}
\newcommand{\tabacro}{Table}
\titlerunning{\gwu}
  \authorrunning{Karathanasis et al. (2022)}

\begin{document} 

\title{\gwu: Python package for creating mock GW samples for different astrophysical populations and cosmological models of binary black holes}

  \author{Christos Karathanasis \inst{\ref{inst1}}\thanks{ckarathanasis@ifae.es}\orcidlink{0000-0002-0642-5507}
  \and 
  Beno\^it Revenu \inst{\ref{inst3},\ref{inst4}}\orcidlink{0000-0002-7629-4805}
  \and
  Suvodip Mukherjee \inst{\ref{inst5}}\thanks{suvodip@tifr.res.in} \orcidlink{0000-0002-3373-5236}
  \and
  Federico Stachurski \inst{\ref{inst2}}\orcidlink{0000-0002-8658-5753}}

  \institute{Institut de Física d’Altes Energies (IFAE), Barcelona Institute of Science and Technology, Barcelona, Spain \label{inst1}
  \and 
  Subatech, CNRS, Institut Mines-Telecom Atlantique, Nantes Universit\'e, France\label{inst3}
  \and
  Université Paris Cité, CNRS, Astroparticule et Cosmologie, France\label{inst4}
  \and
  Department of Astronomy \& Astrophysics, Tata Institute of Fundamental Research, 1, Homi Bhabha Road, Colaba, Mumbai 400005, India\label{inst5}
  \and
     Institute of Gravitational Research (IGR), University of Glasgow, Glasgow, United Kingdom \label{inst2}
    }

    \date{\centering\today }
    
     \abstract{Precision cosmology with gravitational wave (GW) sources requires a deeper understanding of the interplay between GW source population and cosmological parameters governing the dynamics of the Universe. With the swift increase in GW detections, it is necessary to develop a tool for exploring many aspects of cosmology and fundamental physics; this tools allows to simulate GW mock samples for several populations and cosmological models. We have developed a new code called \gwu, which allows us to make a large sample of GW mock events from a broad range of configurations, while varying the cosmology, the merger rate, and the GW source parameters (i.e. mass and spin distributions in particular) for a given network of GW detectors. A large sample of simulated mock GW events will be useful for improving our understanding of the statistical properties of the distribution of GW sources, as long as it is detectable for a given detector noise and an astrophysical and cosmological model. It will also be useful to compare simulated samples with the observed distribution of the GW sources from data and infer the underlying population of the GW source parameters and cosmology. We restricted the cosmology to spatially flat universes, including models with varying dark energy equation of state. The \gwu\ code provides each mock event with a position in the sky and a redshift; these values can be those of random host galaxies coming from an isotropic and homogeneous simulated Universe or a user-supplied galaxy catalog. We used realistic detector configurations of the LIGO and Virgo network of detectors to demonstrate the performance of this code for the latest observation runs and the upcoming observation run.}
     
\keywords{Cosmology: observations--cosmological parameters--Gravitational waves--Galaxies}

\maketitle

 \section{Introduction}
The detection of the gravitational wave (GWs) sources by the LIGO-Virgo-KAGRA collaboration \citep{Harry:2010,Aso:2013,Acernese:2014,LIGOScientific:2014pky,Abbott:2016:gw150914,Martynov:2016fzi,KAGRA:2013pob,Tse:2019,Abbott_2019,KAGRA:2020tym,Abbott_2021,gwtc2.1,https://doi.org/10.48550/arxiv.2111.03606} has opened up an era of GW astronomy capable of bringing a plethora of scientific discoveries in astrophysics, cosmology, and fundamental physics.  With the fourth campaign of data taking O4 that began on May 24, 2023, the expected event rate will rise to some BBHs per day (i.e. more than four times the GW event rate during O3b). Such discoveries will require an understanding of different sources of statistical and systematic uncertainties as well as the interplay between the astrophysical population of GW sources, the cosmological models, the fundamental physics, and searches beyond the standard model \citep{Abbott:2019:population, Abbott:2021:H0,https://doi.org/10.48550/arxiv.2111.03604,https://doi.org/10.48550/arxiv.2111.03634}. To understand these, we need realistic simulations of GW events for different astrophysical populations and cosmological models, including realistic detector sensitivities.

We used \gwu\ to simulate a sample of mock GW sources for different (i) mass, (ii) spin, and (iii) merger rate models, as well as for the flat $\Lambda$ cold dark matter (\lcdm) and flat w$_0$-w$_a$CDM cosmological models. The code assigns host galaxies to GW events in two cases: assuming a constant in comoving volume distribution of galaxies or using a simulated galaxy catalog, such as \mcat~\citep{Fosalba_2014, Crocce_2015,2015MNRAS.448.2987F}. The code also considers different detector sensitivities and observation periods. Then, the \texttt{Bilby}~\citep{Ashton_2019} pipeline is used to generate the actual GW strains corresponding to these GW events and perform parameter estimation to finally obtain mock GW samples.  Having access to these sets of samples allows us to estimate the performances of the analysis algorithms. For one given GW event, its associated samples consist of a list of values obtained by a Markov chain Monte Carlo (MCMC) algorithm of all the parameters describing the binary system; in particular, the luminosity distance of the source, the masses of the two black holes, and their spins.  We applied this simulation code \gwu\ to generate mock sources with the detector sensitivity of the fourth observation run O4 of LIGO-Virgo-KAGRA collaboration and we came up with a forecast for the measurability of the astrophysical population parameters and Hubble constant.
We find that \gwu\ will be useful to generate mock samples for different future detector sensitivities as well as varied astrophysical populations and cosmological parameters.

The paper is structured as follows. In Sect.~\ref{models}, we describe different models included in the current version of \gwu. In Sects.~\ref{code} and~\ref{validation}, we describe the structure of the code and validated different setups.

Finally, in Sect.~\ref{conc}, we discuss the use and future developments of this code. 

\section{Models of cosmology and GW sources astrophysical population} \label{models}

Making inferences on the cosmological results using  binary black holes (BBHs) requires an understanding of the dependencies on the astrophysical source parameters for different techniques such as statistical host identification \citep{Schutz,MacLeod:2007jd,DelPozzo:2012zz,Arabsalmani:2013bj,  PhysRevD.101.122001, Fishbach:2018gjp, Abbott:2021:H0, Soares-Santos:2019irc, Finke:2021aom, Abbott:2020khf, Palmese:2021mjm}, cross-correlation techniques after taking into account GW source population uncertainty \citep{PhysRevD.93.083511,Mukherjee:2018ebj, Mukherjee:2019wcg, Calore_2020,  Mukherjee:2020hyn,Bera:2020jhx, Scelfo:2020jyw, Mukherjee:2020mha,2021ApJ...918...20C,Scelfo:2021fqe,2022MNRAS.511.2782C,Mukherjee:2022afz}, or BBH mass-distribution \citep{Taylor:2011fs,Farr:2019twy, Mastrogiovanni:2021wsd,You:2020wju,Mancarella:2021ecn,2022MNRAS.tmp.2056M,Leyde:2022orh,Ezquiaga:2022zkx}. 

Here, we describe the cosmological models that can be used in \gwu\ and the models describing the population of BBH mergers (merger rates, mass distributions, and spin distributions).

\subsection{Cosmological models}
The code considers only the cosmological models with spatial curvature zero (flat) and for three scenarios namely, \lcdm, w$_0$CDM, which takes into account variation of the dark energy equation of state without redshift dependence, and w$_0$-w$_a$CDM, which takes into account variation of the dark energy equation of state with redshift dependence \citep{Chevallier:2000qy, PhysRevLett.90.091301}. In \tabacro~\ref{tab:cosmo_parameters}, we present these models. 
For a particular choice of the cosmological parameters, the code computes the luminosity distances $D_L(z_m)$ to the merger situated at a redshift~$z_m$ using
\begin{equation}\label{dist}
    D_L(z_m)= \frac{c(1+z_m)}{H_0}\int_0^{z_m} \frac{\dd z}{E(z)},
\end{equation}
with $E^2(z)\equiv {\Omega_m(1+z)^3+ \Omega_{DE}\exp{[\int_0^z\dd\!\ln{(1+z')}(1+\omega(z'))]}}$.

\begin{table*}[!ht]
\centering
\begin{tabular}{ccc}
        \multicolumn{2}{ c}{{\textbf{Cosmology}}}\\
        \hline
        \hline
        \multicolumn{2}{ c}{{Definition of the cosmological parameters}}\\
        \hline
        {\bf Parameter} & \textbf{Description} & \\\hline\hline
        $H_0$ & Hubble constant in km/s/Mpc &\\
        $\Omega_m$ & Present-day matter density of the Universe & \\
        $w(z)= w_0 + w_a \,z/(1+z)$ & Dark energy equation of state and its redshift dependence & \\
        \hline
        \hline
        \\
        \multicolumn{2}{ c}{Cosmological models considered}\\
        \hline
        {\bf Model} & \textbf{Parameters} & \\\hline\hline
        Flat \lcdm:& $H_0$, $\Omega_m$, $w_0=-1$, $w_a=0$\\
        Flat $w_0$CDM:& $H_0$, $\Omega_m$, $w_0$, $w_a=0$\\
        Flat $w_0$-$w_a$CDM :& $H_0$, $\Omega_m$, $w_0$, $w_a$\\
        \hline
        \hline
\end{tabular}
\caption{Description of the cosmological models available and the parameters that govern them.}\label{tab:cosmo_parameters}
\end{table*}

In the part of the code used for assigning the redshift to the GW sources from a galaxy catalog, we can calculate the luminosity distance to the source for different cosmological models. Although this part of the code can be used to infer the cosmological parameters using the luminosity distance, it cannot be used to estimate the variation of cross-correlation between the GW sources and galaxies for different cosmological parameters. To explore the cross-correlation signal, this can only be done in this way for the fiducial cosmological values considered in \mcatv\ simulations. 

\subsection{Merger rate models of the compact objects}
We considered three scenarios for the merger rate of the binary compact objects: (a) constant merger rate, (b) phenomenological merger rate, and (c) delay-time model. The models are described in \tabacro~\ref{tab:models} and their parameters can be seen in \tabacro~\ref{tab:parameters}.

\begin{table*}
\centering
\begin{tabular}{ccc}
        \multicolumn{2}{ c}{{\textbf{Models considered}}}\\
        \hline
        \hline
        \\
        \multicolumn{2}{ c}{{Merger rate models}}\\
        \hline
        {\bf Model} & \textbf{Parameters} & \\\hline\hline
        Redshift-independent :& $R_0,\gamma=0 ,\kappa=0 ,z_p=0$\\
        Redshift-dependent :& $R_0,\gamma ,\kappa ,z_p$\\
        \hline
        \hline
        \\
        \multicolumn{2}{ c}{{Mass models}}\\
        \hline
        {\bf Model} & \textbf{Parameters} & \\\hline\hline
        Truncated power law: & $\alpha,\beta,M_{\rm min},M_{\rm max}$\\
        Broken power law: & $\alpha,\alpha_2,\beta,M_{\rm min},M_{\rm max},\delta_m,b$\\
        Power law+Gaussian: & $\alpha,\beta,M_{\rm min},M_{\rm max},\delta_m,\mu_g,\sigma_g,\lambda_g$\\
        Multi-peak: & $\alpha,\beta,M_{\rm min},M_{\rm max},\delta_m,\mu_g,\sigma_g,\lambda_g,\mu_{\rm{g,low}},\sigma_{\rm{g,low}},\lambda_{\rm{g,low}}$\\
        Redshift-dependent: & All parameters of each model from above+one $\epsilon$ for each parameter.\\
        \hline
        \hline
        \\
        \multicolumn{2}{ c}{{Spin models}}\\
        \hline
        {\bf Model} & \textbf{Parameters} & \\\hline\hline
        Uniform: & --\\
        Gaussian: & $\chi_0,\sigma_0$\\
        Heavy-mass: & $m_{th}$\\
        Correlated: & $\alpha_s,\beta_s,\mu_{\chi,0},\sigma_{\chi,0}$\\
        \hline
        \hline
        \\
        \multicolumn{2}{ c}{{Time delay model}}\\
        \hline
        {\bf Model} & \textbf{Parameters} & \\\hline\hline
        Time-delay: & $R_0,\gamma=2.7,\kappa=2.9,z_p=1.9,\alpha,\beta,M_{\rm min},M_{\rm max},\delta_m,M_{\rm PISN}(Z_*),\sigma_g,\lambda_g,d,t^{min}_d,\alpha_{\rm Z},\gamma_{\rm Z}$\\
        \hline
        \hline
        \\
\end{tabular}
\caption{Description of the population models available.}\label{tab:models}
\end{table*}

\begin{table*}[!ht]
\centering
\begin{tabular}{ccc}
        \multicolumn{2}{ c}{{\textbf{GW source population}}}\\
        \hline
        \hline
        \\
        \multicolumn{2}{ c}{{Definition of the mass distribution parameters}}\\
        \hline
        {\bf Parameter} & \textbf{Description} & \\\hline\hline
        $\alpha$ & Spectral index for the power law of the primary mass distribution & \\
        $\alpha_2$ & Second spectral index for the power law of the primary mass distribution in case & \\
        $\beta$ & Spectral index for the power law of the mass ratio distribution & \\
        $b$ & Point in the mass distribution were the slope changes in $\Msun$ & \\
        $M_{\rm min}$ & Minimum mass of the power law component of the primary mass distribution in $\Msun$ & \\
        $M_{\rm max}$ &  Maximum mass of the power law component of the primary mass distribution in $\Msun$  &\\
        $\mu_{\rm {g}}$ & Mean of the Gaussian component in the primary mass distribution in $\Msun$  &  \\
        $\lambda_{\rm g}$ & Fraction of the model in the Gaussian component &  \\
        $M_{\rm PISN}(Z_*)$ & The value of the PISN mass scale in $\Msun$ units at metallicity $Z_*= 10^{-4}$  &  \\
        $\sigma_{\rm g}$ & Width of the Gaussian component in the primary mass distribution in $\Msun$  & \\
        $\delta_{m}$ & Range of mass tapering at the lower end of the mass distribution in $\Msun$  &\\
        $\lambda_{\rm {g,low}}$ & Fraction of the model in the secondary Gaussian component &  \\
        $\mu_{\rm {g,low}}$ & Mean of the secondary Gaussian component in the primary mass distribution in $\Msun$  &  \\
        $\sigma_{\rm {g,low}}$ & Width of the secondary Gaussian component in the primary mass distribution in $\Msun$  & \\
        $\epsilon_x$ & Redshift-evolution parameter of the $x$ parameter  & \\
        \hline
        \hline
        \\
        \multicolumn{2}{ c}{Definition of the merger rate parameters} & \\
        \hline
        {\bf Parameter} & \textbf{Description} & \\\hline\hline
        $\gamma$ & Power-law index describing the merger rate at low redshift &  \\
        $\kappa$ & Power-law index describing the merger rate at high redshift &\\
        $z_p$ & The redshift where the slope of the merger rate changes  &\\
        $R_0$ & Value of the merger rate at $z=0$ in $\rm Gpc^{-3} \ \rm yr^{-1}$ &\\
        \hline
        \hline
        \\
        \multicolumn{2}{ c}{Definition of the time delay parameters} & \\
        \hline
        {\bf Parameter} & \textbf{Description} & \\\hline\hline
        $d$ & Spectral index for the power law of the delay time distribution &  \\
        $t^{\rm min}_d$ & Minimum time for the power law of the delay time distribution in Gyrs&\\
        $\alpha_{\rm Z}$ & The parameter that captures a weak logarithmic dependence of $M_{\rm PISN}$ on the metallicity  &\\
        $\gamma_{\rm Z}$ & The parameter that captures the redshift dependence of the metallicity  & \\
        \hline
        \hline
        \\
        \multicolumn{2}{ c}{Definition of the spin parameters} & \\
        \hline
        {\bf Parameter} & \textbf{Description} & \\\hline\hline
         $\chi_{0}$ & Mean value of the spin distribution for the Gaussian model  & \\
        $\sigma_{0}$ & Value of the standard deviation of the spin distribution for the Gaussian model  & \\
        $m_{th}$ & Mass threshold value below which spins are set to $0$  & \\
        $\mu_{\chi,{0}}$ & Mean of the Gaussian distribution for $\xeff$  & \\
        $\sigma_{\chi,{0}}$ & Standard deviation of the Gaussian distribution for $\xeff$  & \\
        $\alpha_{s}$ & Mass-ratio correlation parameter for the expectation value  & \\
        $\beta_{s}$ & Mass-ratio correlation parameter for the standard deviation  & \\
        \hline
        \hline
        \\
\end{tabular}
\caption{Description of the parameters that govern the population models.}\label{tab:parameters}
\end{table*}

The first scenario with a constant merger rate model is a model with a single parameter, $R_0$, which corresponds to a scenario with no redshift evolution. The second scenario of the model considers a merger rate varying with the redshift; it uses four phenomenological parameters ($\gamma,~\kappa,~R_0,~z_p$)~\citep{Madau2014}:
\begin{equation}\label{eq:rz-p}
    R(z) = R_0(1+z)^\gamma\,\frac{1+ (1+z_p)^{-(\gamma+\kappa)}}{1+\left(\frac{1+z}{1+z_p}\right)^{\gamma+\kappa}}.
\end{equation}

The third model considered is driven by physics and that models the merger rate in terms of the delay time distribution denoted as $t_d$ \citep{2010ApJ...716..615O,2010MNRAS.402..371B, 2012ApJ...759...52D, Dominik:2014yma, 2016MNRAS.458.2634M, Lamberts:2016txh, 2018MNRAS.474.4997C, Elbert:2017sbr, Eldridge:2018nop, Vitale:2018yhm,  Buisson:2020hoq,Santoliquido:2020axb}.
The time delay is the elapsed time between the formation of the stars (which later will become black holes) and the merging of the black holes. 
Currently, constraints on $t_d$ are weak from individual events \citep{Fishbach:2021mhp, https://doi.org/10.48550/arxiv.2204.13495} and the stochastic GW background \citep{Mukherjee:2021ags}. In the future, by combing the properties of emission line galaxies with GW sources, a better measurement of the delay time is possible \citep{Mukherjee:2021bmw}. The time delay is not the same value for all binary black holes, rather, it follows a distribution of $P_t$:
\begin{equation}\label{eq:t_d}
    P_t(t_d|t_d^{min},t_d^{\rm max},d) \propto
        \begin{cases}
            (t_d)^{-d} & \text{, for $t_d^\text{min}<t_d<t_d^{\rm max}$}\\
            0 & \text{otherwise}\\
        \end{cases}
\end{equation}
For a star to become a black hole only a few Myrs are required, but for a black hole to find another companion, form a binary, and merge a significant amount of time is needed (on the order of Gyrs). This model assumes a star formation rate and different power law forms of the probability distribution of the delay time $P_t(t_d|t_d^{min},t_d^{\rm max},d)$ \citep{2022MNRAS.tmp.2056M,https://doi.org/10.48550/arxiv.2204.13495}. This model has four parameters (d, $t_d^{min}$, $\alpha_Z$, $\gamma_Z$) as described in \tabacro~\ref{tab:parameters} and the corresponding merger rate at redshift, $z$, can be expressed as:
\begin{equation}\label{eq:delay}
    R_\text{TD}(z) = R_0\frac{\int_{z}^{\infty} P_t(t_d|t_d^{min},t_d^{\rm max},d)  R_\text{SFR}(z_f)\der{t}{z_f}\dd z_f}{\int_{0}^{\infty} P_t(t_d|t_d^{min},t_d^{\rm max},d)  R_\text{SFR}(z_f)\der{t}{z_f}\dd z_f}
\end{equation}
 {where $R_\text{SFR}$ is defined as \citep{Madau2014}}
\begin{equation}\label{eq:sfr}
    R_\text{SFR}(z) \propto \frac{(1+z)^{2.7}}{1+[(1+z)/2.9]^{5.6}}.
\end{equation}

For each of these merger rates~$R (z)$, the  number of coalescing events per unit redshift per unit observation time is calculated using
\begin{equation}\label{events}
    \dder{N_{GW}}{z}{t}=  \dfrac{R(z)}{1+z}\der{V_c}{z}(\theta_c),
\end{equation}
where $\der{V_c}{z}(\theta_c)$ is the differential comoving volume at redshift, $z$, which depends on the set of cosmological parameters denoted by~$\theta_c$. For a given observation time, duty cycle, and redshift range, the total number of events can be obtained by integrating the above equation. As an example, using the illustrative values of parameters shown in \tabacro:~\ref{tab:fixedparams}, the number of detectable events is of the order of $150$ ($\pm10\%$) for a total observation time of $3$~years with a network signal-to-noise ratio higher than $(\text{S/N})_\text{thr}=12$. The network (S/N) is defined as the square root of the quadratic sum of the individual (S/N) measured in the interferometers: $\text{(S/N)}=(\sum_i \text{(S/N)}_i^2)^{1/2}$.
\begin{table*}[!ht]
\begin{center}
\begin{tabular}{| c | c || c | c || c |c|}
\hline
\multicolumn{2}{|c||}{Cosmology: flat \lcdm}& \multicolumn{2}{c||}{Merger rate} & \multicolumn{2}{c|}{Mass distributions: Power-law+Gaussian peak} \\
\hline\hline
Parameter & Value & Parameter & Value &Parameter & Value \\
\hline\hline
$H_0$ & $67.8$~\kmsMpc & $\gamma$ & $2.7$ & $\alpha$ & $3.4$ \\
$\Omega_m$ & $0.308$ & $\kappa$ & $2.9$ & $\beta$ & $0.8$\\
$\Omega_\Lambda$ & $0.692$  & $z_p$ & $1.9$ & $\lambda_g$ & $0.04$\\
& & $R_0$ & $20$~\gpcyr & $\mu_g$ & $35~\Msun$\\
& & &  & $\sigma_g$ & $3.9~\Msun$\\
& & & & $\delta_m$ & $4.8~\Msun$\\
& & & & $M_{max}$ & $100~\Msun$\\
& & & & $M_{min}$ & $5~\Msun$\\
\hline
\end{tabular}

\begin{tabular}{| c | c || c | c || c |c|}
\hline
\multicolumn{6}{|c|}{Duty factors}\\
\hline
\multicolumn{2}{|c|}{Detector} & \multicolumn{4}{|c|}{Observation Run}\\
\hline
\multicolumn{2}{|c|}{} & \multicolumn{1}{|c|}{O1} & \multicolumn{1}{|c|}{O2} & \multicolumn{1}{|c|}{O3} & \multicolumn{1}{|c|}{O4}\\
\hline
\multicolumn{2}{|c|}{H}&\multicolumn{1}{|c|}{0.6}&\multicolumn{1}{|c|}{0.6}&\multicolumn{1}{|c|}{0.75}&\multicolumn{1}{|c|}{0.75}\\
\hline
\multicolumn{2}{|c|}{L}&\multicolumn{1}{|c|}{0.5}&\multicolumn{1}{|c|}{0.6}&\multicolumn{1}{|c|}{0.75}&\multicolumn{1}{|c|}{0.75}\\
\hline
\multicolumn{2}{|c|}{V}&\multicolumn{1}{|c|}{-}&\multicolumn{1}{|c|}{-}&\multicolumn{1}{|c|}{0.75}&\multicolumn{1}{|c|}{0.75}\\
\hline
\end{tabular}
\caption{Tables show the fiducial value of the parameters and duty cycle used in the analysis.}\label{tab:fixedparams}
\end{center}
\end{table*}

\subsection{Mass models} The GW signal strongly depends on the individual masses $m_1$ and $m_2$ (with $m_1\geqslant m_2$) of the two black holes forming the compact system. The joint distribution $p(m_1,m_2)$ and the marginalized $p(m_1)$ and $p(m_2)$ depend on the black hole formation channels. We broadly considered two scenarios of phenomenological mass models, namely the redshift-independent mass models and redshift-dependent mass models. The mass models are described in \tabacro~\ref{tab:models} and their associated parameters in \tabacro~\ref{tab:parameters}. The mathematical description is given in Appendix~\ref{app:1}.

\subsubsection{Redshift-independent mass models}
For the redshift-independent mass models, we include scenarios with only (a) a truncated power-law model without smoothing (i.e. there is a sharp cut-off at low mass values), (b) a broken power-law model with smoothing, (c) a power-law+Gaussian peak model with smoothing (d) a model with power-law+multipeak Gaussian model with smoothing. All parameters are listed in \tabacro~\ref{tab:parameters}. 

The redshift independent models are motivated by the current studies of the rates and population of the GW sources from nearly ninety GW sources discovered by the LVK collaboration \citep{Abbott_2021,https://doi.org/10.48550/arxiv.2111.03634}. For these mass models, the code creates a mock sample of the GW sources which follows the underlying distribution of both the companion masses $m_1$ and $m_2$ according to \citep{Talbot_2018,Abbott2020b,Virgo:2021bbr,LIGOScientific:2021psn}:
\begin{align}\label{eq:masses}
    P_s (m_1,m_2 | \Phi_m) = &p(m_1| \Phi_m)\,p(m_2 |m_1, \Phi_m)\nonumber\\ & \times S(m_{1}|\delta_m,M_{\rm min})\, S(m_{2}|\delta_m,M_{\rm min}),
\end{align}
where $\Phi_m$ refers to all the population parameters not related to cosmology and $p(m_2 |m_1, \Phi_m)$ is the distribution of $m_2$, conditioned on the constraint $m_1\geqslant m_2$, and $S$ is the function used to smooth the lower end of the mass distributions, taken as a sigmoid. For the model without smoothing, $S=1$. \citep{https://doi.org/10.48550/arxiv.2111.03634} is considered.  {The functional form of each of these terms is provided in Appendix \ref{app:1}.}

\subsubsection{Redshift-dependent mass models}
The redshift dependence of the mass distribution of BBHs can arise due to the evolution of the stellar properties in the Universe \citep{Spera:2017fyx,2019ApJ...887...53F, Renzo:2020rzx, 2022MNRAS.tmp.2056M}. Two different scenarios of the redshift-dependent mass model are considered in the analysis, namely (a) a phenomenological model which considers the redshift evolution of the mass distribution, and (b) a mixing-mass model which considers the dependence of the binary masses on the stellar metallicity and delay time distribution. 

For the phenomenological model with a mass distribution dependent on the redshift, we considered a linear dependence modeled as $\mu(z)= \mu_0+ \epsilon_\mu z$, where $\mu_0$ denotes the expectation value of the mass of the redshift independent case (discussed previously) and $\epsilon_\mu$ is the slope of the linear evolution of this expectation value. This model is an extension of the redshift-independent model with one additional parameter for each of the redshift-independent mass model parameters. As a result, there are~$2N$ parameters, if the redshift independent model has~$N$ parameters.  {In Appendix \ref{app:1}, we describe different phenomenological models with redshift-dependent mass distribution.}

\begin{figure*}[!ht]
    \centering
    \includegraphics[scale=.4]{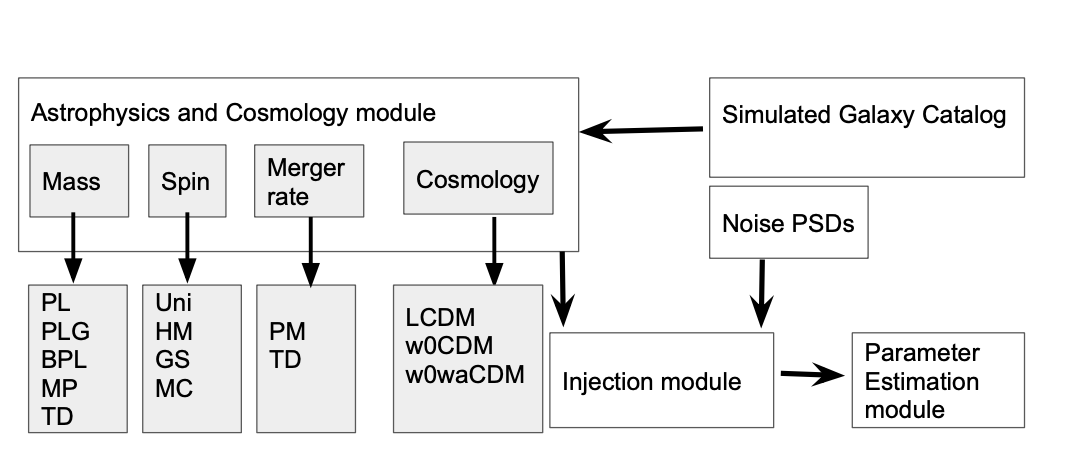}
    \includegraphics[scale=.4]{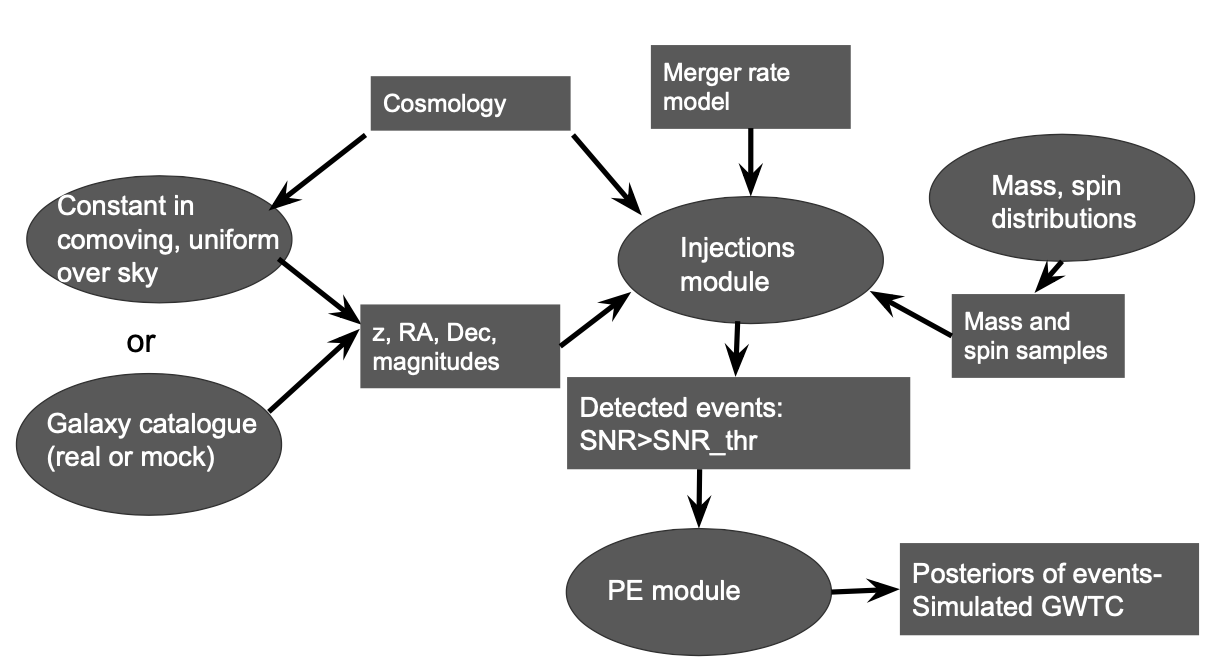}
    \caption{\gwu\ flow chart. (Top) Description of the modules of the \gwu{} code. Available models for the masses: Power-Law (PL), Power-Law with Gaussian (PLG), Broken Power-Law (BPL), Multi Peak (MP), and the time-delay (TD) model. For the spin: Uniform (Uni), Heavy-Mass (HM), Gaussian Spin (GS), and Mass-spin Correlated (MC). For the merger rate: phenomenological model (PM), time-delay (TD). For the cosmology: \lcdm\, w$_0$CDM, w$_0$-w$_a$CDM). (Bottom) In the plot, we describe the logical structure of the \gwu{} code and how the injection samples are produced.}
    \label{fig:code}
\end{figure*}

In the mixing-mass model \citep{2022MNRAS.tmp.2056M,https://doi.org/10.48550/arxiv.2204.13495}, the mass distributions of the compact objects get a redshift dependence due to a  metallicity evolution as a function of redshift and a non-zero value of the delay time distribution. Due to the non-zero value of the delay time (as $t_d^{min} >0$), black holes forming at different cosmic times originate from different parent stars with different metallicities and consequently can have different masses. Consequently, when the black holes form at different cosmic times, they merge after a different amount of delay time and the observed BH mass distribution gets a redshift dependence. This redshift-dependent mass distribution $P[m(z_m)]$ can be written in terms of the window function $W_{t_d}[m(z_m)]$ and the source frame mass distribution $P_s[m(z_m)]$ of the compact objects as 
\begin{equation}\label{mass}
    P[m(z_m)] = P_s[m(z_m)]W_{t_d}[m(z_m)],
\end{equation}
where the window function can be written as 
\begin{equation}
   W_{t_d}[m(z_m)] = N \int_{z_m}^{\infty}P_t(t_d|t_d^{min},t_d^{\rm max},d) \der{t}{z_f}\,W[m(z_{f})]\,\dd z_{f},
\end{equation}
with $P_t$ defined in \eq~\ref{eq:t_d} and $N$ is a normalization constant. Further details about this model can be found in \citep{2022MNRAS.tmp.2056M,https://doi.org/10.48550/arxiv.2204.13495}. 

\subsection{Spin models}
The GW strain also depends on the individual spins (amplitude and orientation) of both compact objects. The study of the spin of GW sources can shed light on the astrophysical population and the formation channel of these sources \citep{Mandel:2009nx,Kushnir:2016zee,Rodriguez:2016vmx,Stevenson:2017dlk, OShaughnessy:2017eks,Farr:2017uvj, Gerosa:2018wbw,Fuller:2019sxi,Bavera:2020inc,Biscoveanu:2020are, Biscoveanu:2021nvg,Biscoveanu:2022qac}.  
The spin of a black hole is defined as $\vec{S}=S\vec{\hat{S}}$, where $|\vec{\hat{S}}|=1$ and ${S=Gm^2\chi/c}$, with $\chi$ being a dimensionless scalar parameter in the range of $\chi\in[0;1]$. Denoting the total angular momentum of the binary system as $\vec{J}$, the tilt angles of the spin vectors with respect to $\vec{J}$ are defined as $\cos\theta_i=\vec{\hat{S}}_i\cdot\vec{J}/J$. Aligned spins correspond to the cases $\theta_1=\theta_2=0$ or $\theta_1=\theta_2=\pi$.
In practice, GW measurements are sensitive to the following combination of the spins, which defines the effective spin $\xeff$ of the binary system:
\beq
\xeff = \frac{m_1\,\chi_1\cos\theta_1+m_2\,\chi_2\cos\theta_2}{m_1+m_2}=\frac{\chi_1\cos\theta_1+q\,\chi_2\cos\theta_2}{1+q},
\eeq
where $q$ is the mass-ratio $q=m_2/m_1$($q\in[0;1]$) and $\xeff$ takes values in the range of $\xeff\in[-1;1]$. Individual spin components are not very well constrained with the current interferometers. Thus, we deduce the individual values from $\xeff$ and not the other way.  {These individual spin values (amplitude and geometry) are needed in the computation of the theoretical GW signal with \texttt{Bilby}}; in Sect.~\ref{sec:zindep}, we describe how we proceed for this.

We are considering two families of scenarios for the spin distribution of the binary compact objects, namely the redshift-independent and the redshift-dependent scenarios. The spin models are described in \tabacro~\ref{tab:models} and the details of their parameters are summarised in~\tabacro~\ref{tab:parameters}.  {We have described all the models considered in this analysis in Appendix \ref{app:3}.}

\subsubsection{Redshift-independent spin models}\label{sec:zindep}
The redshift-independent case is considered for two classes of models  {namely: (i) mass spin uncorrelated and (ii) mass-spin correlated model. For the mass spin uncorrelated model, we considered two cases (a) the uniform spin model and (b) the Gaussian spin model. For the mass-spin correlated model, we considered two cases, (a) the heavy-mass model and (iv) mass-spin correlated Gaussian model. We describe all the models in Appendix \ref{app:3}.}

\textit{Mass spin uncorrelated model:} In the uniform spin model and Gaussian spin distribution model, for each GW event, a random $\xeff$ spin value is drawn from a uniform distribution $\mathcal{U}[-1,1]$ or from a Gaussian distribution, $\mathcal{N}(\mu_{\xeff},\sigma_{\xeff})$, with an expectation value of $\mu_{\xeff}$ and a standard deviation of $\sigma_{\xeff}$. Then, the individual parameters $(\chi_i,\theta_i)$ are computed given $\xeff$ but there is a strong degeneracy: many tuples of $(\chi_1,\theta_1,\chi_2,\theta_2)$ for a given $q$ correspond to the same $\xeff$ so, here we present one way to get such a tuple. If the spins are requested to be aligned then the computation of $(\chi_i,\theta_i)$ is performed as follows: set first both angles $\theta_i$ to $0$ or $\pi$ for a positive and negative value of $\xeff$, respectively. We then draw both values of $\chi_i$ as follows:
\begin{gather*}
    \chi_2 \sim \mathcal{U}[a,b]\\
    \text{with}~a=\max(0,((1+q)|\xeff|-1)/q),\\
    \text{and}~b=\min(1,(1+(1+q)|\xeff|)/q)\\
    \text{then}~\chi_1 = (1+q)|\xeff|-q\chi_2.
\end{gather*}
If the spin directions are requested as random, then we draw the values following:
\begin{gather*}
    \chi_2\cos\theta_2 \sim \mathcal{U}[a;b]\\
    \text{with}~a=\max(-1,((1+q)\xeff-1)/q),\\
    \text{and}~b=\min(1,(1+(1+q)\xeff)/q)\\
    \text{then}~\chi_1\cos\theta_1 = (1+q)\xeff-q\chi_2\cos\theta_2.
\end{gather*}
The degeneracy between individual spins $\chi_i$ and tilt angles $\theta_i$ is broken with:
\begin{gather*}
    \chi_2=|\chi_2\cos\theta_2|^x~\text{with}~x\sim\mathcal{U}[0;1]\\
    \text{and finally}~\cos\theta_2=\chi_2\cos\theta_2/\chi_2\\
    \chi_1=|\chi_1\cos\theta_1|^y~\text{with}~y\sim\mathcal{U}[0;1]\\
    \text{and finally}~\cos\theta_1=\chi_2\cos\theta_1/\chi_1.
\end{gather*}

The second class of spin models that are independent of redshift allow for a correlation to be found between the individual masses and the spin values. We have considered two sub-classes of redshift independent spin models under: \textit{(i)} objects with heavier masses have higher individual spin $\chi_i$, \textit{(ii)} the mass and spin of the black holes are (anti)correlated. In the first case, we choose a free threshold parameter of the mass values $m_{\rm th}$, above which the individual spin values are not zero and below which the spin is set to zero. In the second case, we include a correlation between the mass ratio and spin values as motivated in \citep{Callister:2021fpo}. In this model, $\xeff$ follows a Gaussian distribution 
\begin{equation}\label{spin}
p(\chi_\text{eff}|q,\mu_{\chi,0},\sigma_{\chi,0},\alpha_s,\beta_s)
 \propto \exp
\left[-
\frac{ \left(\chi_\text{eff}-\mu_\chi(\mu_{\chi,0},\alpha_s,q)\right)^2}
{ 2\sigma_{\chi}^2(\sigma_{\chi,0},\beta_s,q)}\right],
\end{equation}
with an expectation value $\mu_\chi=\mu_{\chi,0}+\alpha_s(q-1/2)$ and a variance, $\sigma_{\chi}^2(\sigma_{\chi,0},\beta_s,q)$, such that $\log_{10} \sigma_\chi(\sigma_{\chi,0},\beta_s,q)=\log_{10} \sigma_{\chi,0}+\beta_s(q-1/2)$. For $\alpha_s=\beta_s=0$, the model is mass-independent and~$\xeff$ follows the Gaussian case discussed at the beginning of this section, $\xeff\sim\mathcal{N}(\mu_{\chi,0},\sigma_{\chi,0})$.

\subsubsection{Redshift-dependent spin models}\label{sec:spinz}
We considered one scenario of the redshift-dependent model, where the individual spin amplitude distribution is considered as Gaussian with a redshift-dependent expectation value and standard deviation $P(\chi_{i}|z)= \mathcal{N}(\mu_{\chi,0}(z), \sigma_{\chi,0}(z))$. We have described the model in the Appendix~\ref{app:spin-rd}.
We consider a linear evolution of both the expectation value $\mu_{\chi,0}(z)$ and
standard deviation $\sigma_{\chi,0}(z)$ of the spin amplitude.
The spin orientations can be taken as aligned ($\theta_1=\theta_2=0$ or $\theta_1=\theta_2=\pi$ randomly) or isotropically distributed.

\section{Code description}\label{code}
\gwu\ is a code to make simulated GW transient catalogs (GWTCs) by assigning GW events to galaxies. The galaxies are taken from a galaxy catalog (or without the usage of a galaxy catalog, simply creating a redshift distribution of galaxies assuming they are distributed uniformly in comoving volume, namely, without any spatial clustering) for different astrophysical models of the binary compact objects and different cosmological models. The basic structure of the code and a flow chart showing the working principle of the code are presented in Fig.~\ref{fig:code}. We describe the structure of the code below.

The code is composed of five different modules, (i) astrophysical and cosmological module, (ii) galaxy catalog module, (iii) detector noise module, (iv) injection module, and (v) parameter estimation module. Under the astrophysical module, we have four sub-modules which include different models of the merger rate, source frame mass population, spin distribution, and cosmological models. Each of these modules contains different models which are discussed in Sect.~\ref{models}. The galaxy catalog module includes functions to handle simulated galaxy catalogs which can be used to populate the GW sources into galaxies. In the next version of this code, scenarios with a real galaxy catalog will also be included. The third module is the detector one which includes the noise power spectral density (PSD) for different detectors such as Hanford (H), Livingston (L), Virgo  (V) and different LVK observation runs such as O1, O2, O3, and O4.
These three modules are called by the injection module. The injection module is used to take into consideration all the aforementioned and simulate a GW strain of all GW sources for different choices of the astrophysical population models, cosmological models, detector noise, duty cycles, and observation time. The different GW strains are injected into the detector noise and the code calculates the (S/N) of all simulated events. 

\subsection{Simulated GWTC without a galaxy catalog}

 {In the absence of a galaxy catalog, we first have to create a custom fake universe.
In this first step, the user needs to specify the density of galaxies, the cosmological parameters, the Schechter function parameters and the redshift range. Galaxies are sampled using a uniform in comobile volume distribution. The universe can be used afterwards to assign galaxy hosts to GW events.}

Then, in the injection module, the code uses a particular model of the merger rate and cosmology (specified by the user, it must be the same cosmology than the one used to create the fake universe) and calculates the number of GW events (using \eq~\eqref{events}) up to a redshift of $z_{\rm max}$. The value of $z_{\rm max}$ can be decided by the user. For each of the GW events, the code draw random values from the source-frame mass and spin distributions. The galaxy host is randomly chosen in agreement with the merger rate and using a weight that can be uniform or proportional to the galaxy luminosity; the redshift of the event and its sky position are taken to be those of the host galaxy. In this case, the galaxies are assumed to follow a uniform comoving volume distribution and they are distributed uniformly over the sky. 
From the chosen redshifts, the luminosity distances to the GW sources are calculated using the selected cosmological model. 

The final step of this process is to estimate the matched filtering (S/N) for a chosen network of detectors accordingly to the duty factors and observation time, using the corresponding noise PSDs (which are available in the detector module). The GW events having a (S/N) greater than the detection threshold $\text{(S/N)}_\text{thr}$ chosen by a user make up the detected events and then undergo parameter estimation using the publicly available code~\texttt{Bilby}. 

\begin{figure}
    \includegraphics[scale=0.5]{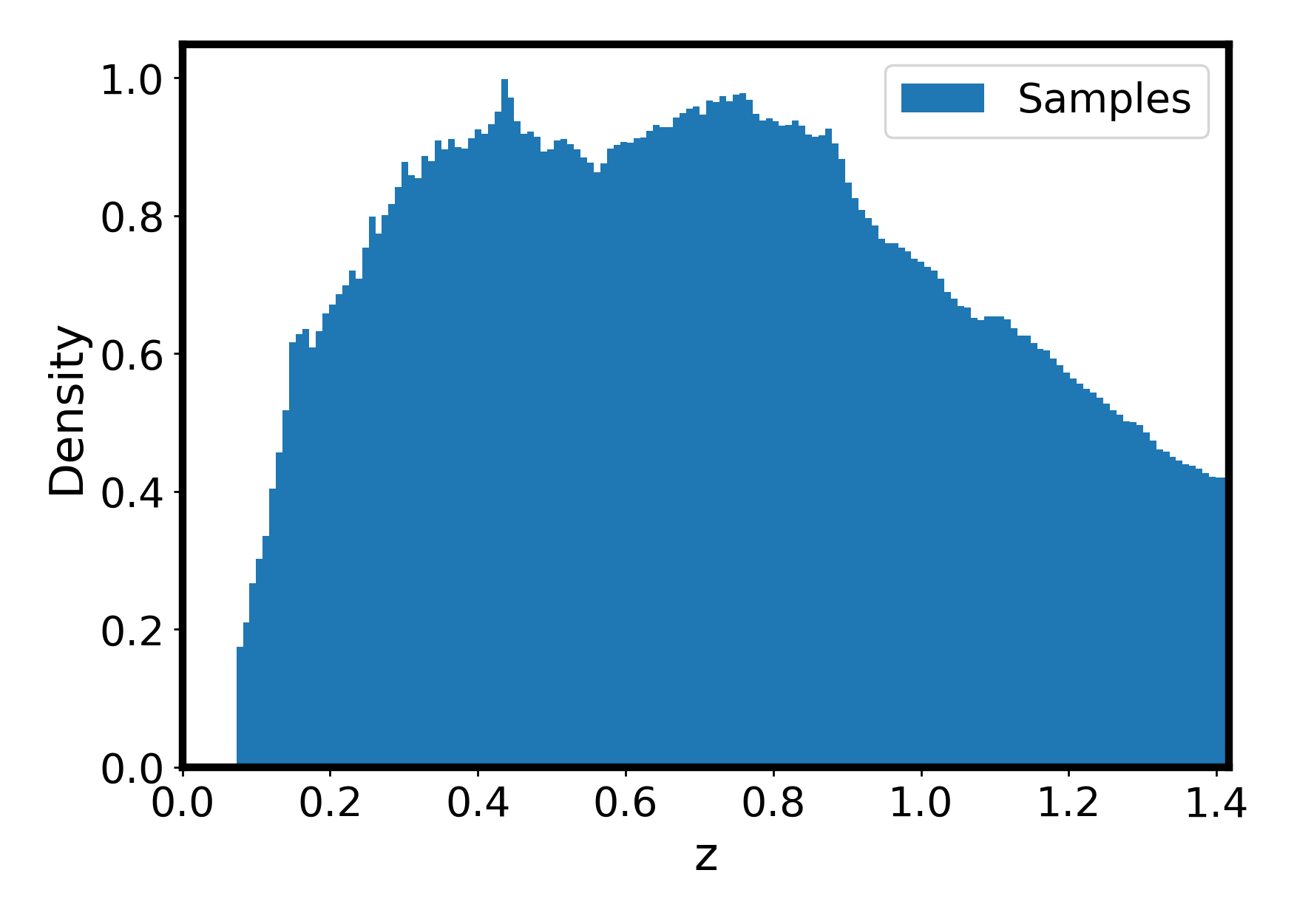}
    \caption{True redshift distribution of galaxies in the simulated catalog \mcat. }
    \label{fig:MICE_z}
\end{figure}

\begin{figure}
    \includegraphics[scale=0.4]{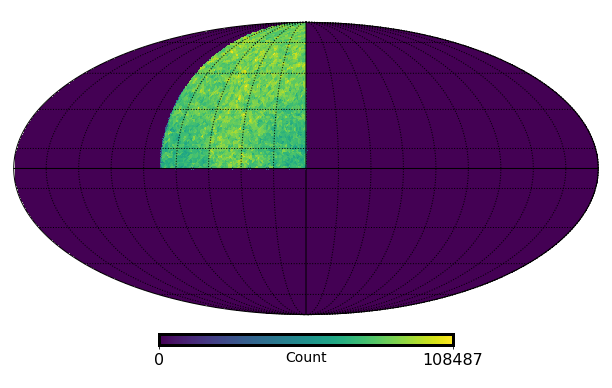}
    \caption{Molleview projection of bin count per pixel in \mcat. }
    \label{fig:MICE_sky}
\end{figure}

\subsection{Simulated GWTC using a galaxy catalog}
The primary difference between using or not using a galaxy catalog is the underlying spatial distribution of galaxies that GW sources are going to be assigned to. The galaxies in the simulated galaxy catalog contain spatial clustering, whereas when the galaxies are chosen without a galaxy catalog (fake universe case), they do not feature spatial clustering as they are uniformly distributed in comobile volume. The former can be used to explore the cross-correlation technique whereas for the second one, cross-correlation cannot be used.

In the case when a galaxy catalog is used, the code loads from the catalog the galaxies that are assigned with luminosities in a certain band (chosen by a user). The underlying distribution of the redshifts of galaxies (in this case) is the one from the galaxy catalog, before we apply any merger rate model. The distribution of galaxies in the sky depends on the galaxy catalog. Most of the galaxy catalogs are only covering a small fraction of the sky. In this case, GW events are going to originate from that part only. Full sky galaxy catalogs, like GLADE+~\citep{10.1093/mnras/stac1443}, can be used to avoid this limitation. In the future, this module will be developed further to include different formation channels of GW sources and host galaxy properties.

At present, the code cannot take into account the galaxy catalog's incompleteness. This will be added in the future where \gwu $ $ will be able to account for galaxies that host GW events but are not present in the galaxy catalog.

 {Below, we describe a mock galaxy catalog that is incorporated in the \gwu{} code.} We have used the mock galaxy catalog \emph{\texttt{MICE}-Grand Challenge Galaxy and Halo Light-cone Catalog} (\mcat) \citep{Fosalba_2014,Crocce_2015,2015MNRAS.448.2987F} developed using an $N$-body simulation containing about 70 billion dark matter particles in a $3~\text{Gpc}/h^{3}$ comoving volume. The dark matter halos and galaxies generated from this span a sky area of 5000 $\text{deg}^{2}$ and up to a redshift of $z = 1.4$, with no simulation box repetition. The galaxy catalog is made with the following values of the cosmological parameters for a flat-\lcdm\ model $\Omega_{m}$ = 0.25; $\sigma_{8} = 0.8$; $n_{s}$ = 0.95; $\Omega_{b}$ = 0.044; $ \Omega_{\Lambda}$ = 0.75 and $h = 0.7$\footnote{The catalog is available for download from the \emph{CosmoHub} website. \citep{https://doi.org/10.48550/arxiv.2003.03217}}.

The redshift distribution in \mcat\ is shown in Fig.~\ref{fig:MICE_z}. The luminosities and magnitudes of the simulated galaxies are provided in the same bands than actual cosmological surveys: DES \cite{DES_survey}, Euclid \cite{Euclid_survey}, COSMOS \cite{COSMOS_Survey}, SDSS \cite{SDSS_Survey}, VISTA \cite{VISTA_Survey}. We applied an evolutionary correction to the apparent magnitudes as suggested in \mcat.\footnote{\mcat's "readme" document~(\url{https://www.dropbox.com/s/0ffa8e7463n8h1q/README_MICECAT_v2.0_for_new_CosmoHub.pdf?dl=0})}
\begin{equation}\label{mag_corr}
m_\text{evol} = m_\text{true} - 0.8\,  \bigg(\arctan\big(1.5\, z\big)- 0.1489\bigg),
\end{equation}
where $m_\text{true}$ is the observed magnitude in the catalog, $z$ is the true redshift and $m_\text{evol}$ is the evolved luminosity in redshift. We use $m_{\text{evol}}$ as the observed apparent magnitude, which is used in the luminosity weighting calculation: the user can ask for the probability for a galaxy to host a merger to be proportional to its absolute luminosity. The galaxy catalog also follows the observed luminosity functions, color distributions, and clustering
as a function of luminosity and color at low-$z$.

 {From \mcat, GW host galaxies are chosen following the GW merger rate model (discussed in the previous section). At each redshift, for a given merger model and observation time, we calculate the number of GW sources $N_{\rm GW}$ using \eq~\eqref{events}. Then, we chose $N_{\rm GW}$ galaxies from the catalog for every redshift and consider these as the hosts of the GW sources. The galaxies chosen as the hosts of the GW sources can also be chosen based on the luminosity of the galaxies at different user-specified bands. We can use a luminosity weighting to choose the GW sources as follows:}
\begin{equation}\label{eq:lum_wei}
    w_{i} = 10^{-0.4\,M_{i}},
\end{equation}
 {where $M_{i}$ is the absolute magnitude for the $i$th band, for a specific survey. The value of $M$ depends on different wavelength bands in which galaxies are observed and in this code, a user can choose a specific survey and an observation band. To compute the absolute magnitude of the selected galaxy, one can either use the given absolute magnitudes from \mcat\ or compute them directly from the true correct apparent magnitudes from \eq~\ref{mag_corr} for the $i$th band as}

\begin{equation}\label{eq:abs_mag}
    M_{i} = m_{\text{evol}, i} - 5\log_{10} \Bigg(\frac{d_{L}(z,H_{0})}{\text{Mpc}} \Bigg) - 25,
\end{equation}

 {where $d_{L}(z,H_{0})$ is the luminosity distance of the galaxy, as a function of the measured redshift and the Hubble constant, in units of Mpc.}

In the following discussion, we explain how \mcat\ can be used in \gwu\ to assign galaxies from the catalog to GWTC events. We start by reading the catalog file and by selecting the data columns that we are interested in. We select the sky location in equatorial coordinates (RA and Dec), the true redshift of the galaxies, and the observed magnitude, $m_{\text{evol}}$, for a specific band. 

Since \mcat~is a simulated galaxy catalog, no redshift uncertainty is present. However, \gwu\ can add redshift uncertainties by taking a certain percentage (user given) of the true redshift and using this as the 1$\sigma$ standard deviation. It can also apply masking to the catalog, where it counts the number of galaxies per pixel over the entire sky, and when this number is smaller than a certain threshold (user given), it treats the pixel as empty. The masking of the catalog over the sky is shown in Fig.~\ref{fig:MICE_sky}, as it can be seen, it covers one-eighth of the sky. This masking function will be more useful in future works that adopt real galaxy catalogs, such as GLADE+, thereby introducing a way to measure the angular completeness in each respective catalog.

\begin{figure*}[!ht]
    \centering
    \includegraphics[scale=.45]{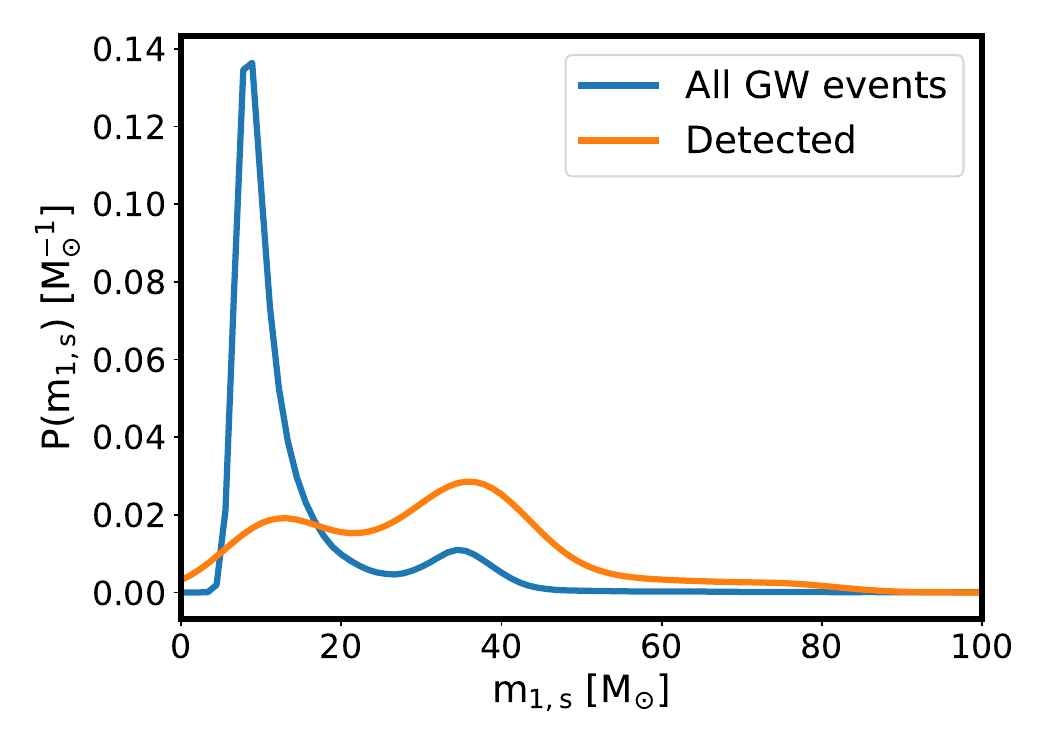}
    \includegraphics[scale=.45]{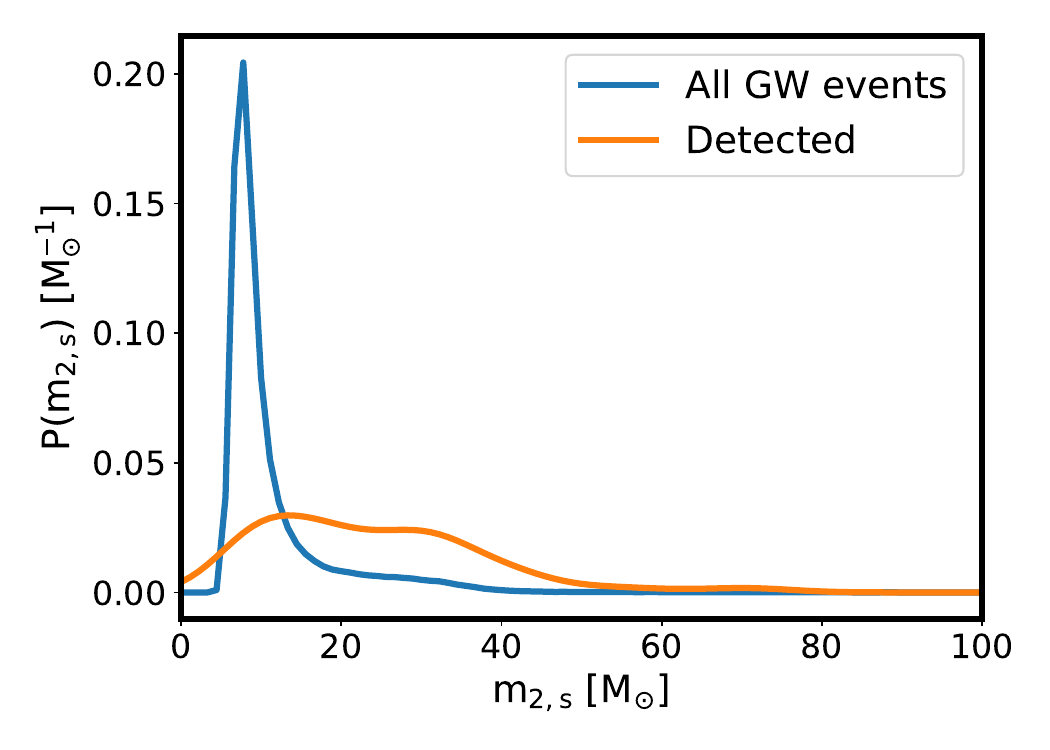}
    \begin{center}
        \includegraphics[scale=.45]{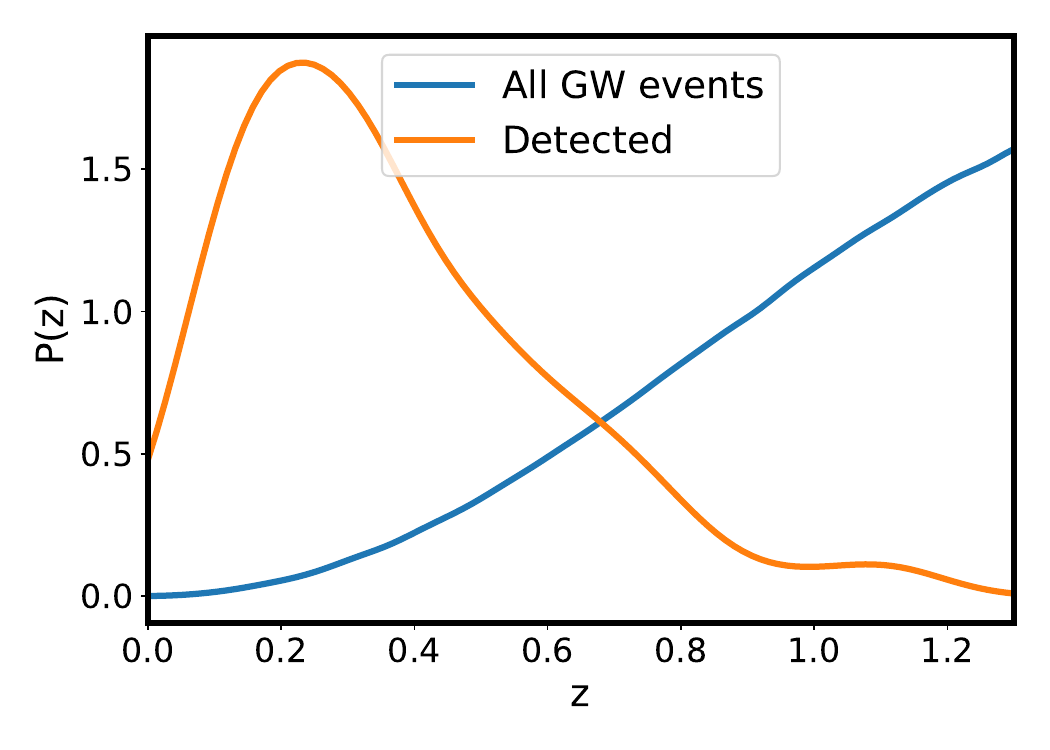}
    \end{center}
    \caption{True values of the primary component masses (top left) and lighter component masses (top right) distributions (analytic expression) in the source frame of the detected events for a simulation of O1-, O2-, O3-, and O4-like 3 observational years. The redshifts (bottom) distribution of the events is also shown. The network (S/N) threshold was set to (S/N)$_\text{thr}=12$. In the same plots, the distribution (KDE estimation) of all the simulated (detected and not detected) events can be seen. The population, cosmological, merger rate, and detector parameters are given in \tabacro~\ref{tab:fixedparams}.}
    \label{fig:detectd_vs_all}
\end{figure*}

\section{Code validation}\label{validation}
The validation of the code is done in two steps. The first one (i) consists in checking that the distributions are correctly sampled. This implies we check whether the mock catalog of GW sources matches the expected population of the GW sources. Then (ii) we produced a mock GW catalog and, by using codes that can infer cosmological parameters from GW data, we try to retrieve the injected cosmological parameters.

\textit{Comparison of the GW mock samples with the expected population:} Starting from the redshift-independent mass models, various distributions and corresponding samples can be seen in Fig.~\ref{fig:mass_samples}. The analytical expressions of the distributions are plotted as dashed orange curves whereas the drawn samples are histogrammed in blue. The comoving volume and the merger rate distributions as functions of redshift can be seen in Fig.~\ref{fig:redshift_samples}, as well as samples drawn from them. The code can also assign spins to events sampling from the models described in \tabacro~\ref{tab:models}. For the results presented from the simulation here, we kept all the spins fixed to zero but an example of sampling of spins for the mass correlated model can be seen in Fig.~\ref{fig:spin}.  {Thus, for all relevant parameters such as masses, merger rate, and spin, \gwu{} can generate mock samples of the desired distribution.}

\textit{ {Inference of the injected parameters for O4 Observation run of LIGO-Virgo-KAGRA:}} \gwu\ computes the number of expected BBH events for certain values of the merger rate model parameters and an observational time selected by the user. Using a uniform comoving volume distribution as the galaxies' redshift distribution, we drew hosts of GW events following the merger rate model presented in \eq~\ref{eq:rz-p}, with $\gamma=2.7, \ \kappa=2.9, \ z_p=1.9, \ R_0=20 \ \rm{Gpc^{-3} \ yr^{-1}}$. We simulated 3 years of O1-, O2-, O3- and O4-like observation run for a realistic duty cycle with the power law+Gaussian distribution for $m_1$ that is shown in Fig.~\ref{fig:mass_samples} (bottom-left). The duty factors for O1, O2, and O3  were similar to the real ones for the actual LVK observational runs, whereas for O4, we selected values equal to the ones of O3, which can be seen in \tabacro~\ref{tab:fixedparams}. 

The total number of mock BBH events is more than 200 thousand (calculated using $z_{max}=5$). Of those, using a network (S/N) threshold (S/N)$_\text{thr}=12$, only 164 are detected (recalling that the network (S/N) is the (S/N) obtained when combining data from all interferometers).  {In Fig. \ref{fig:detectd_vs_all}, we show the kernel density estimate (KDE) of the masses and reshift (all events and detected ones only). This makes the distribution appears smooth. The comparison of the normalized distributions of all the simulated events and detected ones denotes the selection fraction as a function of masses and distance. As heavier sources are more likely to be detected than the lighter ones up to a higher luminosity distance, the distribution of the injected sources and the detected ones differs. This shows the expected distributions of the detected sources with O1+O2+O3+O4 observation with the duty cycle as mentioned in \tabacro~\ref{tab:fixedparams}.}

The detected events that pass the (S/N)$_\text{thr}$ undergo parameter estimation using \texttt{Bilby}, to obtain the posteriors of their parameters, namely individual masses ($m_1$ and $m_2$) in detector frame, luminosity distance ($D_L$), source position (RA and Dec), and the inclination angle ($\theta_{JN}$) for a single event. 
Having a set of posterior samples at our disposal allows us to use a population and/or cosmology estimation code and retrieve the injected values. We used the python package \texttt{IcaroGW} which can estimate jointly the population parameters of BBHs and the cosmological parameters \citep{Mastrogiovanni_2021}. We used a power-law+Gaussian peak distribution for the primary mass model and the flat-\lcdm\ for cosmology.  {In \texttt{IcaroGW} we fix $\Omega_m$=0.3 and allow only $H_0$ to vary. The posteriors obtained on the black hole population parameters and the Hubble constant for O1+O2+O3+O4 configuration are shown in Fig.~\ref{fig:icaro_results} along with the injected values (black dashed lines). This joint estimation of 13 parameters including the parameters describing the merger rate parameters ($R_0, \gamma, \kappa, z_p$), black hole mass distribution ($\alpha, \beta, \rm M_{\rm max},\rm M_{\rm min}, \mu_g, \sigma_g, \lambda_g, \delta_m$) and cosmology ($H_0$) presents a forecast for measuring these quantities from the upcoming observation run. The posterior distribution of the population parameters agrees within $95\%$ C.L. with the injected values for most of the parameters. From this joint estimation, the injected value of $H_0$ is aptly measured from these sources. Among the merger rate-related parameters, we find that the local merger rate $R_0$ and the power-law index, $\gamma$, describing the redshift evolution of the low redshift merger rate of the binary black holes are well constrained. However, the parameters such as $\kappa$ and $z_p$ which describe the black hole merger rate at high redshift are not well constrained, as with the LIGO-Virgo-KAGRA O4 sensitivity, most of the detected binaries lie below redshift $z=1$. Among the parameters describing the mass distribution, the parameters such as the power-law index of the masses, $\alpha$, $\beta$, and maximum mass, $\rm M_{\rm max}$, are better constrained than the other parameters describing the mass population. This is because these parameters impact a wide range of masses and several detected sources contribute to their estimation. However, some parameters such as $\rm M_{\rm min}, \delta_m$ which captures the minimum mass and smoothing scale of the mass distribution at the lower masses can be poorly constrained as the lower end of the mass distribution is difficult to measure because the sensitivity of detecting lighter sources is smaller. Similarly, at the higher mass end, the parameters such as $\mu_g, \sigma_g$, and $\lambda_g$ which capture a localized Gaussian feature in the mass distribution can be constrained by only those sources which are detected around this mass range. With the O4 sensitivity and a duty cycle given in \tabacro~\ref{tab:fixedparams}, only a few sources have been adequately measured, which contributes to constraining these parameters. As a result, due to statistical fluctuations, there is a large scatter in the inference of these parameters. The distribution of the sources in the simulated set agrees well with fiducial values, as shown in Fig. \ref{fig:mass_samples} for the mass distribution mismatch. The mismatch in the estimation of the fiducial value shown in Fig. \ref{fig:icaro_results} is due to the statistical fluctuation.

 {To further explore how the mass distribution gets impacted by the statistical fluctuations, we obtain the distribution of masses $m_{1,s}$ and $m_{2,s}$ in the source frame of the detected events with an (S/N) $>12$ for different noise realizations (having the same noise power spectrum) with $3$ years and $20$ years of observation time, respectively. In Fig.~\ref{fig:comparison} we show the results for fiducial population, cosmological, population, and detector configuration mentioned in \tabacro~\ref{tab:fixedparams}. The plot shows that with $3$ years of detectable events, different noise realizations cause a larger statistical fluctuation in the probability distribution of masses. However, as the number of sources increases with $20$ years of observation time, the statistical fluctuation reduces, and the mass distribution of the detectable events converges to the expected distribution of the selected events as shown by orange lines in Fig. \ref{fig:detectd_vs_all}.} 

\begin{figure*}
    \centering
    \includegraphics[scale=0.25]{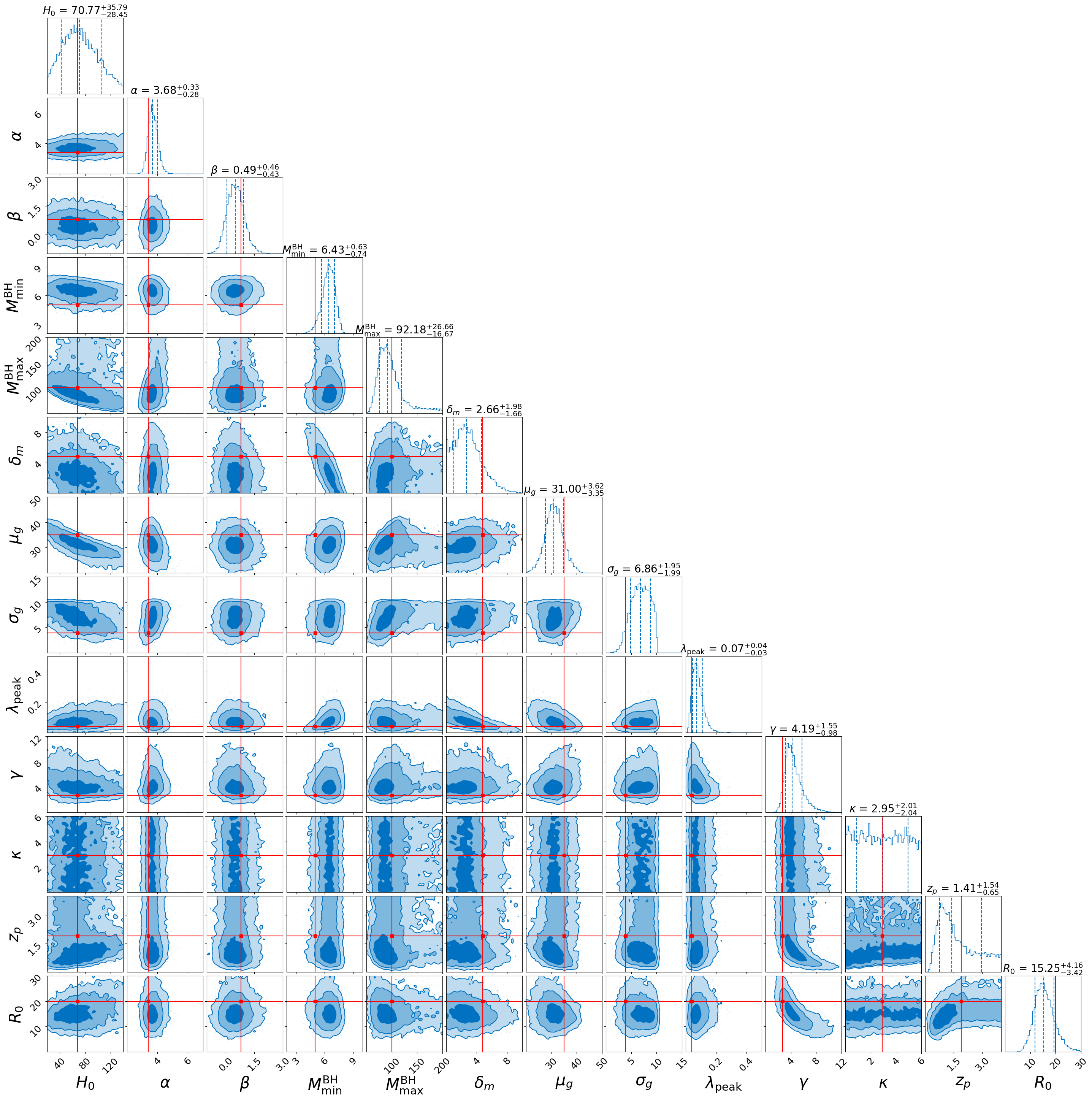}
    \caption{Population and cosmological parameter posteriors obtained from an O1-, O2-, O3-, and O4-like simulation using the cosmological inference python package \texttt{IcaroGW}. The red lines indicate the injected values of the parameters.}
    \label{fig:icaro_results}
\end{figure*}

\begin{figure*}
    \centering
    \includegraphics[scale=.45]{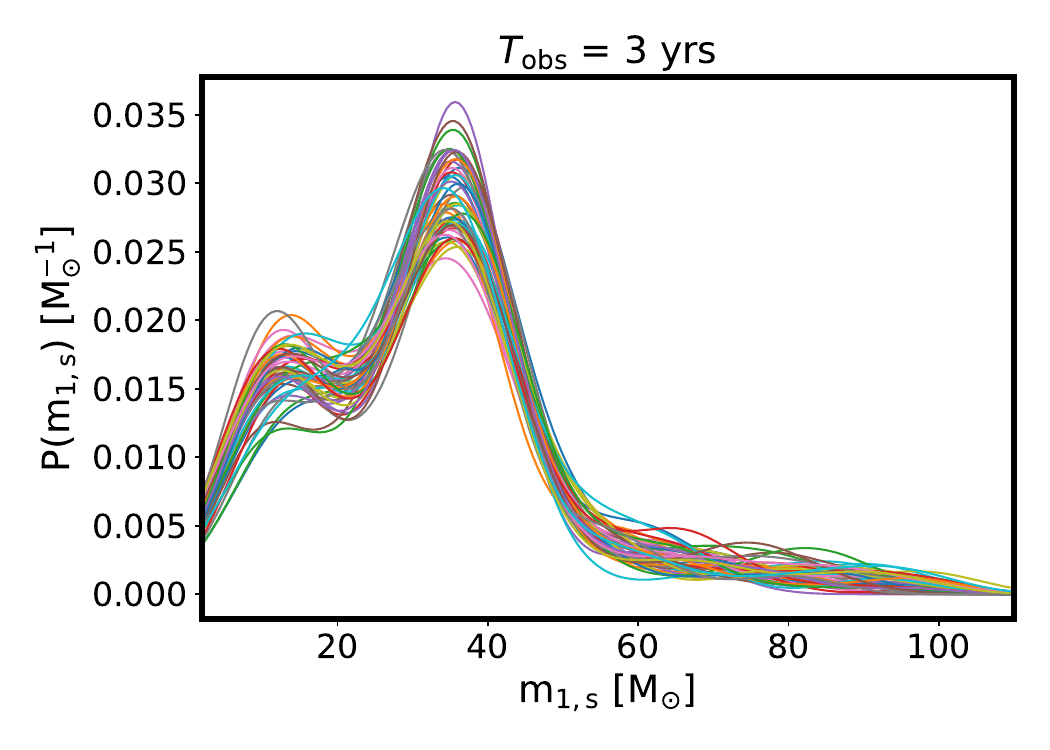}
    \includegraphics[scale=.45]{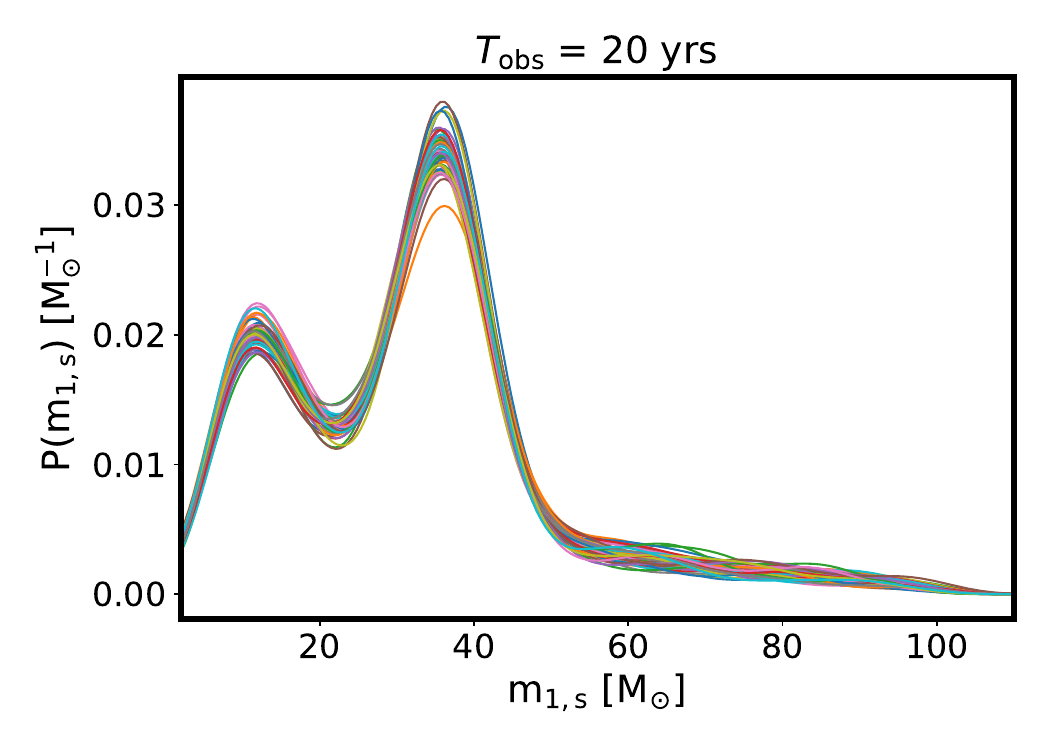}
\includegraphics[scale=.45]{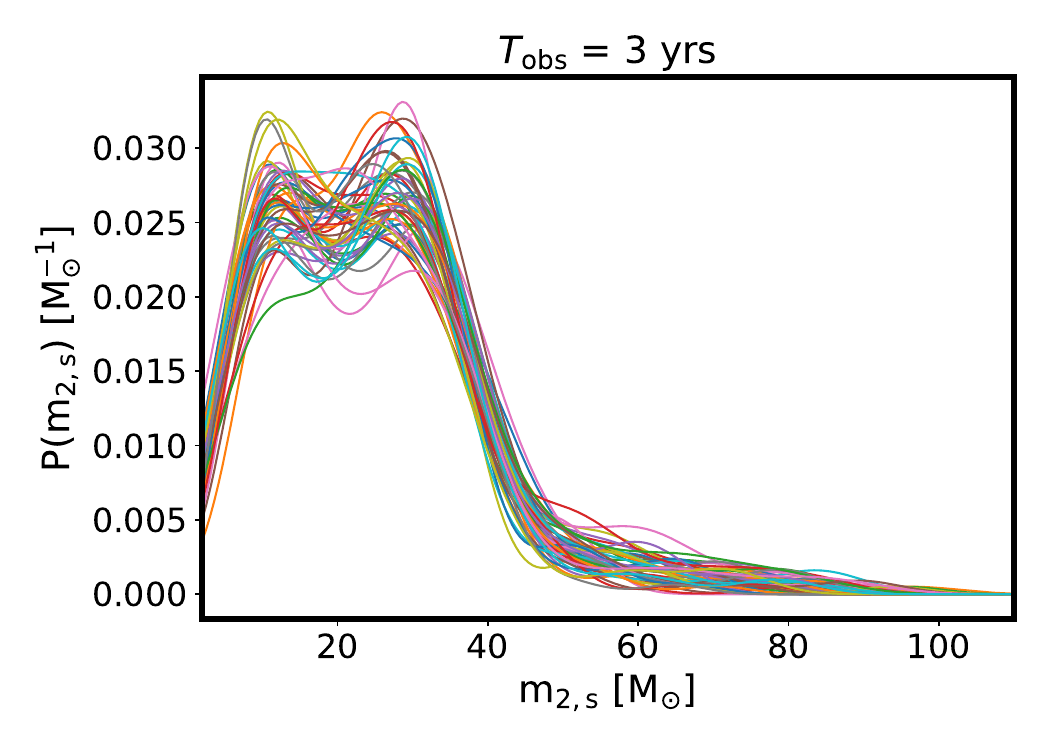}
    \includegraphics[scale=.45]{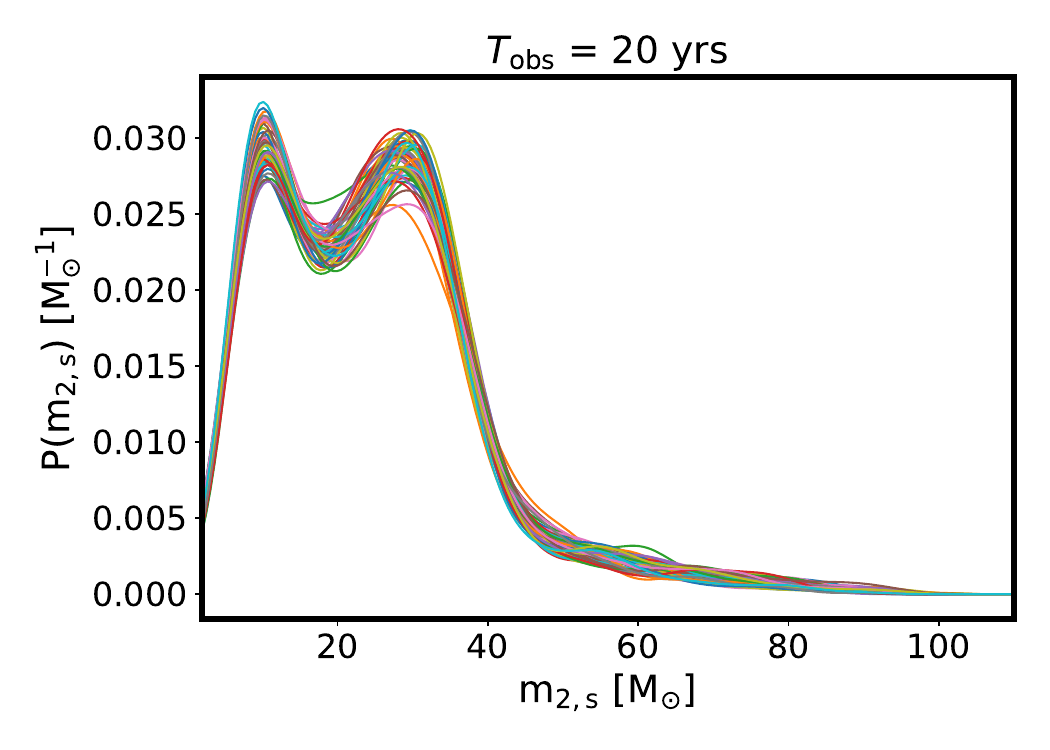}
    \caption{Various $m_1$ (top) and $m_2$ (bottom) distributions of the detected events with $\rm (S/N)>12$ for $T_{\rm obs}=3$ years (left) and $T_{\rm obs}=20$ years (right) produced for different random seeds. Raising the total observational time leads to more events being detected and the distributions being better defined. The population, cosmological, merger rate, and detector parameters used are given in \tabacro~\ref{tab:fixedparams}  {and the corresponding distributions are represented by the orange lines of Fig.~\ref{fig:detectd_vs_all}}.}
    \label{fig:comparison}
\end{figure*}

\section{Conclusions}\label{conc}
We present a new \gwu\ package to simulate GW sources for different cosmological and GW population models. This code can incorporate five scenarios of GW mass models, four scenarios of GW spin distributions, two scenarios of the merger rate, and three models of the cosmological parameters, as described in Fig.~\ref{fig:code}. The code will be useful to simulate realistic mock samples of the GW sources for different networks of GW detectors. Currently, only second-generation detectors are being used for simulations but in the future, next-generation detectors such as the Cosmic Explorer \citep{Reitze:2019iox,Hall:2020daa} and Einstein Telescope \citep{Punturo:2010zz} will also be included. For a chosen model of the GW source population parameters and cosmological parameters, the code generates samples of binary compact objects for a fixed amount of observation time, duty cycle, and configurations of the detector network. The code can also make use of a simulated galaxy catalog \mcatv\ for assigning the mock GW samples to the redshift and sky position of the host galaxies, possibly using a luminosity weighting. In the future, a revised version will be made to include a real galaxy catalog such as GLADE+ \citep{10.1093/mnras/stac1443} and \citet{Duncan:2022wpi}. The \gwu\ code is also wrapped up with the code \texttt{Bilby} for parameter estimation of the GW source parameters. This code will be useful for carrying out simulations of GW sources and understand the interplay between GW source population and cosmology. This code can be used for the currently ongoing network of detectors and also for the next-generation GW detectors.   

\gwu\ is publicly available online.\footnote{\url{https://git.ligo.org/benoit.revenu/gwsim}}

\section*{Acknowledgements}
The authors thank Simone Mastrogiovanni for his helpful review of the manuscript as a part of LVK P\&P. 
C. Karathanasis is partially supported
by the Spanish MINECO under the grants SEV-2016-0588 and PGC2018-101858-B-I00, some of which include ERDF funds from the European Union. IFAE is partially funded by the CERCA program of the Generalitat de Catalunya. The work of S. Mukherjee is a part of the $\langle \texttt{data|theory}\rangle$ \texttt{Universe-Lab} which is supported by the TIFR and the Department of Atomic Energy, Government of India. 
This analysis is carried out at the computing facility of the LSC cluster. We acknowledge the use of data from the MICE simulations, publicly available at \url{http://www.ice.cat/mice}. This work has made use of CosmoHub. CosmoHub has been developed by the Port d'Informació Científica (PIC), maintained through a collaboration of the Institut de Física d'Altes Energies (IFAE) and the Centro de Investigaciones Energéticas, Medioambientales y Tecnológicas (CIEMAT) and the Institute of Space Sciences (CSIC and IEEC), and was partially funded by the "Plan Estatal de Investigación Científica y Técnica y de Innovación" program of the Spanish government. We acknowledge the use of the following packages in this work: Astropy \citep{2013A&A...558A..33A, 2018AJ....156..123A}, \texttt{Bilby} \citep{Ashton_2019}, Giant-Triangle-Confusogram \citep{Bocquet2016}, IPython \citep{PER-GRA:2007}, Matplotlib \citep{Hunter:2007},  NumPy \citep{2011CSE....13b..22V}, and SciPy \citep{scipy}. The authors are grateful for computational resources provided by the LIGO Laboratory and supported by National Science Foundation Grants PHY-0757058 and PHY-0823459. This research has made use of data or software obtained from the Gravitational Wave Open Science Center (gw-openscience.org), a service of LIGO Laboratory, the LIGO Scientific Collaboration, the Virgo Collaboration, and KAGRA. LIGO Laboratory and Advanced LIGO are funded by the United States National Science Foundation (NSF) as well as the Science and Technology Facilities Council (STFC) of the United Kingdom, the Max-Planck-Society (MPS), and the State of Niedersachsen/Germany for support of the construction of Advanced LIGO and construction and operation of the GEO600 detector. Additional support for Advanced LIGO was provided by the Australian Research Council. Virgo is funded, through the European Gravitational Observatory (EGO), the French Centre National de Recherche Scientifique (CNRS), the Italian Istituto Nazionale di Fisica Nucleare (INFN), and the Dutch Nikhef, with contributions by institutions from Belgium, Germany, Greece, Hungary, Ireland, Japan, Monaco, Poland, Portugal, Spain. The construction and operation of KAGRA are funded by the Ministry of Education, Culture, Sports, Science and Technology (MEXT), Japan Society for the Promotion of Science (JSPS), National Research Foundation (NRF), and Ministry of Science and ICT (MSIT) in Korea, Academia Sinica (AS) and the Ministry of Science and Technology (MoST) in Taiwan. This material is based upon work supported by NSF's LIGO Laboratory which is a major facility fully funded by the National Science Foundation.

\bibliographystyle{aa} 
\bibliography{paper_draft}

\appendix

\section{ {Mass distributions}}\label{app:1}
Here, we describe some of the available mass distributions.
The implemented mass models are considered for two classes, (i) redshift-independent mass models and (ii) redshift-dependent mass models. 
\subsection{Redshift-independent mass models}\label{app:zindepmass}
\subsubsection{Truncated power law}\label{app:1a}
 In this case the set of mass distribution parameters is $\Lambda_{m_1}=\{M_{min},M_{max},\alpha\}$ and the distribution is given by:
\begin{equation}
    p_{PL}(m_1|\Lambda_{\rm m_1})\propto
    \begin{cases}
            m_1^{-\alpha} & M_{min}<m_1<M_{max}\\
            0 & \text{otherwise}
        \end{cases}.
\end{equation}
\subsubsection{Broken power law}\label{app:1b}
In this case, the distribution is a combination of two power-law distributions with indexes $a_1,a_2$ joined together at a breaking point defined by $m_{break}=M_{min}+b(M_{max}-M_{min})$. We also introduced a tapering factor $\delta_m$ that smooths the lower end of the distribution. In this case we have $\Lambda_{m_1}=\{M_{min},M_{max},\alpha_1,\alpha_2,b,\delta_m\}$ and:
\begin{equation}
    \begin{split}
        p_{BPL}(m_1|\Lambda_{m_1})\propto S(m_1|M_{min},\delta_m) \times \\
        \begin{cases}
                m_1^{-\alpha_1} & M_{min}<m_1<m_{break}\\
                m_1^{-\alpha_2} & m_{break}<m_1<M_{max}\\
                0 & \text{ otherwise}
            \end{cases},
    \end{split}
\end{equation}
where:
\begin{equation}\label{eq:smooth}
    S(m_1|M_{min},\delta)=
    \begin{cases}
            0 & m_1<M_{min}\\
            f(m_1-M_{min},\delta_m) &  M_{min}<m_1<M_{min}+\delta_m\\
            1 & m_1>M_{min}+\delta_m
        \end{cases},
\end{equation}
is the smoothing function defined by:\\ $f(m,\delta_m)=\left(1+\exp\left(\dfrac{\delta}{m}+\dfrac{\delta}{m-\delta}\right)\right)^{-1}$. 

\subsubsection{Power law Gaussian}\label{app:1c}
This model is a power-law distribution superimposed with a Gaussian peak distribution with mean $\mu_g$ and width $\sigma_g$. Here we need an additional parameter that controls the fraction of the Gaussian peak with respect to the power law $\lambda_g$. We again used the tapering factor and in this case, we have:\\ $\Lambda_{m_1}=\{M_{min}, M_{max},\alpha,\mu_g,\sigma_g,\lambda_g,\delta_m\}$ and:
\begin{equation}
    \begin{split}
        & p_{PLG}(m_1|\Lambda_{\rm m_1})\propto S(m_1|M_{min} ,\delta_m) \times \\
        & \begin{cases}
            (1-\lambda_g)p_{PL}(m_1|M_{max},M_{min},\alpha)\\+\lambda_g G(m_1|\mu_g,\sigma_g)& M_{min}<m_1<M_{max}\\
            0 & \text{ otherwise}
        \end{cases},
    \end{split}
\end{equation}
where 
\begin{equation}\label{eq:a5}
    G(m_1|\mu_g,\sigma_g)=\dfrac{1}{\sigma_g\sqrt{2\pi}}\exp\left(-\dfrac{(m_1-\mu_g)^2}{2\sigma_g^2}\right).
\end{equation}
\subsubsection{Multi-peak}\label{app:1d}
The multi-peak model describes a power-law distribution superimposed with two Gaussian peaks. The distribution is given by:
\begin{equation}
    \begin{split}
        & p_{MP}(m_1|\Lambda_{\rm m_1})\propto 
         S(m_1|M_{min} ,\delta_m)  \times \\
        & \begin{cases}
            (1-\lambda_g)p_{PL}(m_1|M_{max},M_{min},\alpha)+\\\lambda_g(1-\lambda_{g,low}) G(m_1|\mu_g,\sigma_g)+\\\lambda_g\lambda_{g,low} G(m_1|\mu_{g,low},\sigma_{g,low})& M_{min}<m_1<M_{max}\\
            0 & \text{ otherwise}
        \end{cases},
    \end{split}
\end{equation}
where the parameter $\lambda_g$ is the fraction of events in the two Gaussian components, while the $\lambda_{g, low}$ is the fraction of events in the lower Gaussian component.

Finally, the probability density function of the mass of the lightest black hole ($m_2$) is defined by:
\begin{equation}
    p(m_2|m_1)\propto
    \begin{cases}
            m_2^{\beta} & M_{min}<m_2<m_1\\
            0 & \text{otherwise}
        \end{cases}.
\end{equation}
It is a conditional power law in the mass interval $[M_{min};m_1]$, conditioned on the value of $m_1$.

\subsection{Redshift-dependent mass models}\label{app:1e}
The phenomenological redshift-dependent mass model allows us to make the simulation of the redshift evolution of the astrophysical population of compact objects and enables us to study of how well such effects can be measured from GW observations. We have considered the phenomenological redshift-dependent model for all the models described in Appendix \ref{app:1a}-\ref{app:1d}, namely: (i) the truncated power-law, (ii) broken power law, (iii) power-law Gaussian, and (iv) multi-peak model.  We allowed for a linear variation in the model parameters with redshift for all these models. We describe below the redshift dependence of the parameters considered in the analysis. 

\subsubsection{Redshift-dependent truncated power law model}
In this case, we can allow for a variation in the parameters $M_{\rm min}, M_{\rm max}$ and $\alpha$ as
\begin{align}
\begin{split}
    M_{\rm min}(z)=& M_{\rm min} + \epsilon_{M_{\rm min}} z,\\
     M_{\rm max}(z)=& M_{\rm max} + \epsilon_{M_{\rm max}} z,\\
      \alpha(z)=& \alpha + \epsilon_{\alpha} z.
      \end{split}
\end{align}
The probability distribution is described by 
\begin{equation}
     p_{PL}(m_1|\Lambda^z_{\rm m_1})_\propto
    \begin{cases}
            m_1^{-\alpha(z)} & M_{min}(z)<m_1<M_{max}(z)\\
            0 & \text{otherwise}
        \end{cases}
\end{equation}
\subsubsection{Redshift-dependent broken power law model}
For the redshift-dependent broken power-law model, we allow for a linear evolution of the model parameters with redshift. The evolution of the model parameters can be expressed as:
\begin{align}
\begin{split}
    M_{\rm min}(z)=& M_{\rm min} + \epsilon_{M_{\rm min}} z,\\
     M_{\rm max}(z)=& M_{\rm max} + \epsilon_{M_{\rm max}} z,\\
      \alpha_1(z)=& \alpha_1 + \epsilon_{\alpha_1} z,\\
       \alpha_2(z)=& \alpha_2 + \epsilon_{\alpha_2} z,\\
        b(z)=& b + \epsilon_{b} z,\\
        \delta_{m}(z)=& \delta_{m} + \epsilon_{\delta_{m}} z.
        \end{split}
\end{align}
The probability distribution of the broken power-law model is 
\begin{equation}
    \begin{split}
        p_{BPL}(m_1|\Lambda^z_{\rm m_1})\propto S(m_1|M_{min}(z),\delta_m(z)) \times \\
        \begin{cases}
                m_1^{-\alpha_1(z)} & M_{min}(z)<m_1<m_{break}(z)\\
                m_1^{-\alpha_2(z)} & m_{break}(z)<m_1<M_{max}(z)\\
                0 & \text{ otherwise,}
            \end{cases},
    \end{split}
\end{equation}
where $m_{\rm break}(z)= M_{min}(z)+b(z)(M_{max}(z)-M_{min}(z))$
The functional form of $S(m_1|M_{min}(z),\delta_m(z))$ is shown in \eq~\eqref{eq:smooth}. 

\subsubsection{Redshift-dependent power-law Gaussian model}
The redshift-dependent power-law Gaussian model considers a linear variation in the model parameters with redshift as:
\begin{align}
\begin{split}
    M_{\rm min}(z)=& M_{\rm min} + \epsilon_{M_{\rm min}} z,\\
     M_{\rm max}(z)=& M_{\rm max} + \epsilon_{M_{\rm max}} z,\\
      \alpha(z)=& \alpha + \epsilon_{\alpha} z,\\
       \mu_g(z)=& mu_g + \epsilon_{mu_g} z,\\
        \sigma(z)=& \sigma + \epsilon_{\sigma} z,\\
        \lambda_g(z)=& \lambda_g + \epsilon_{\lambda_g} z,\\
        \delta_{m}(z)=& \delta_{m} + \epsilon_{\delta_{m}} z.
        \end{split}
\end{align}
The probability distribution of the mass distribution is 
\begin{equation}
    \begin{split}
        & p_{PLG}(m_1|\Lambda^z_{\rm m_1})\propto S(m_1|M_{min} (z),\delta_m(z)) \times \\
        & \begin{cases}
            (1-\lambda_g(z))p_{PL}(m_1|M_{max}(z),M_{min}(z),\alpha(z))\\+\lambda_g(z) G(m_1|\mu_g(z),\sigma_g(z))& \hspace{-0.7cm}M_{min}(z)<m_1<M_{max}(z),\\
            0 & \text{ otherwise,}
        \end{cases}
    \end{split}
\end{equation}
where the functional form of $G(m_1|\mu_g(z),\sigma_g(z))$ is:
\begin{equation}\label{eq:a5z}
    G(m_1|\mu_g(z),\sigma_g(z))=\dfrac{1}{\sigma_g(z)\sqrt{2\pi}}\exp\left(-\dfrac{(m_1-\mu_g(z))^2}{2\sigma_g^2(z)}\right).
\end{equation}

\subsubsection{Redshift-dependent multi-peak model}
The redshift-dependent multi-peak model allows for the redshift evolution of the peaks in the mass distribution. We consider the redshift evolution of the parameters as:
\begin{align}
\begin{split}
    M_{\rm min}(z)=& M_{\rm min} + \epsilon_{M_{\rm min}} z,\\
     M_{\rm max}(z)=& M_{\rm max} + \epsilon_{M_{\rm max}} z,\\
      \alpha(z)=& \alpha + \epsilon_{\alpha} z,\\
       \mu_g(z)=& mu_g + \epsilon_{mu_g} z,\\
        \sigma(z)=& \sigma + \epsilon_{\sigma} z,\\
        \lambda_g(z)=& \lambda_g + \epsilon_{\lambda_g} z,\\
        \mu_{g, low}(z)=& mu_{g, low} + \epsilon_{mu_{g, low}} z,\\
        \sigma_{g, low}(z)=& \sigma_{g, low} + \epsilon_{\sigma_{g, low}} z,\\
        \lambda_{g, low}(z)=& \lambda_{g, low} + \epsilon_{\lambda_{g, low}} z,\\
        \delta_{m}(z)=& \delta_{m} + \epsilon_{\delta_{m}} z,
        \end{split}
\end{align}
The probability distribution of the mass distribution is 
\begin{equation}
    \begin{split}A
               & p_{MP}(m_1|\Lambda^z_{\rm m_1})\propto 
         S(m_1|M_{min}(z) ,\delta_m(z))  \times \\
        & \begin{cases}
            (1-\lambda_g(z))p_{PL}(m_1|M_{max}(z),M_{min}(z),\alpha(z))+\\\lambda_g(z)(1-\lambda_{g,low}(z)) G(m_1|\mu_g(z),\sigma_g(z))+\\\lambda_g(z)\lambda_{g,low}(z) G(m_1|\mu_{g,low}(z),\sigma_{g,low}(z))& \hspace{-0.7cm}M_{min}(z)<m_1<M_{max}(z)\\
            0 & \text{ otherwise,}
        \end{cases}
    \end{split}
\end{equation}
where the functional form of $G(m_1|\mu_g(z),\sigma_g(z))$ given in \eq~\eqref{eq:a5z}. 

\subsubsection{Mixing-mass model}
To capture the mass distribution of BBHs at a given merging redshift, $z_m$, we consider the following:
\begin{multline}
    p_{TD}(m_1|z_m,\Lambda_{m_1},\Lambda_{d},\Lambda_{nuis}) = p(m_1|z_m,\Lambda_{m_1})W_{t_d}(m_1;z_m) \\
    \times\,S(m_1|M_{min} ,\delta_m),
\end{multline}
where the window function $W_{t_d}(m;z_m)$ is calculated using:
\begin{equation}\label{eq:window_function}
    W_{t_d}(m;z_m) = N \int_{z_m}^{\infty}P_t(t_d|t_d^{\rm min},t_d^{\rm max},d)\der{t}{z_f}W(m;z_{f})\,\dd z_{f},
\end{equation}
where $N$ is a normalization factor, $P_t$ is the delay time distribution, $W[m(z)]$ is a Heaviside step function $W(m;z_f)=\Theta(M_{\rm PISN}(z_f) - m)$, and $z_{f}$  is the redshift of the formation of a BH. The $W_{t_d}$ function brings a breaking point $M_{\rm break}$ at the mass distribution. The delay time distribution is taken to be a simple power-law function of the delay time $t_d$:
\begin{equation}
    P_t(t_d|t_d^{\rm min},t_d^{\rm max},d) \propto
        \begin{cases}
            (t_d)^{-d} & \text{for~~~~} t_d^{\rm min}<t_d<t_d^{\rm max},\\
            0 & \text{otherwise,}\\
        \end{cases}
\end{equation}
and the delay time is given by $t_d=t_m-t_f$, with the notation $t_m=t(z_m), t_f=t(z_f)$ to be the time of merger and time of formation, respectively. 
The distribution of $m_1$ in the source frame $p(m_1|z_m,\Phi_{m_1})$ is considered to be given by a power-law distribution superpositioned with the distribution of a Gaussian peak:
\begin{multline}
    p(m_1 |z_m, \Lambda_{m_1}) = (1 -\lambda_g)P(m_1|M_{\rm min},M_{\rm max},-\alpha)+\\
    \lambda_{g} G(m_1|M_{\rm break}(z_m),\sigma_g),
\end{multline}
where\\
$\Lambda_{m_1}=\{M_{\rm min},M_{\rm max},\alpha,\lambda_{g},M_{\rm break}(z_m), \sigma_{g}\}$, $G(m_1|M_{\rm break}(z_m),\sigma_g)$ (as defined in \eq~\ref{eq:a5})
and $P(m_1|M_{\rm min},M_{\rm max},-\alpha)$ is a power-law distribution with slope $-\alpha$ between $M_{\rm min}$ and $M_{\rm max}$.

In Fig.~\ref{fig:mass_samples}, the histogram of the samples drawn from the input distributions is shown in blue and the analytical curves are shown as orange dashed curves.

\begin{figure*}[!ht]
    \centering
    \includegraphics[scale=.45]{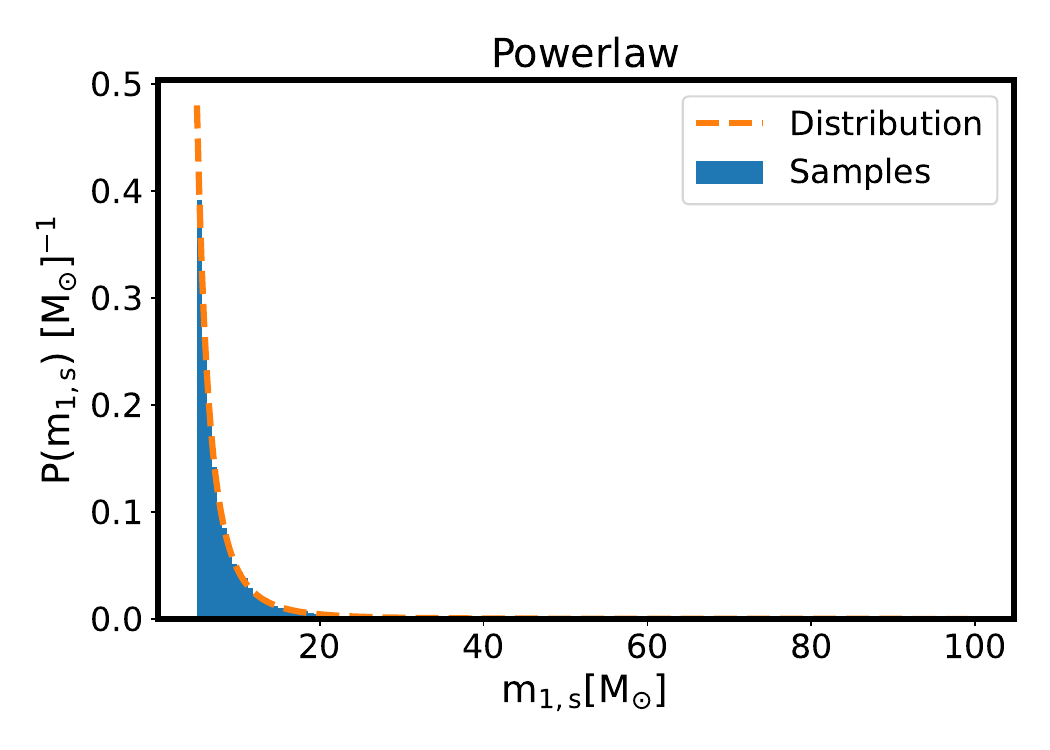}
    \includegraphics[scale=.45]{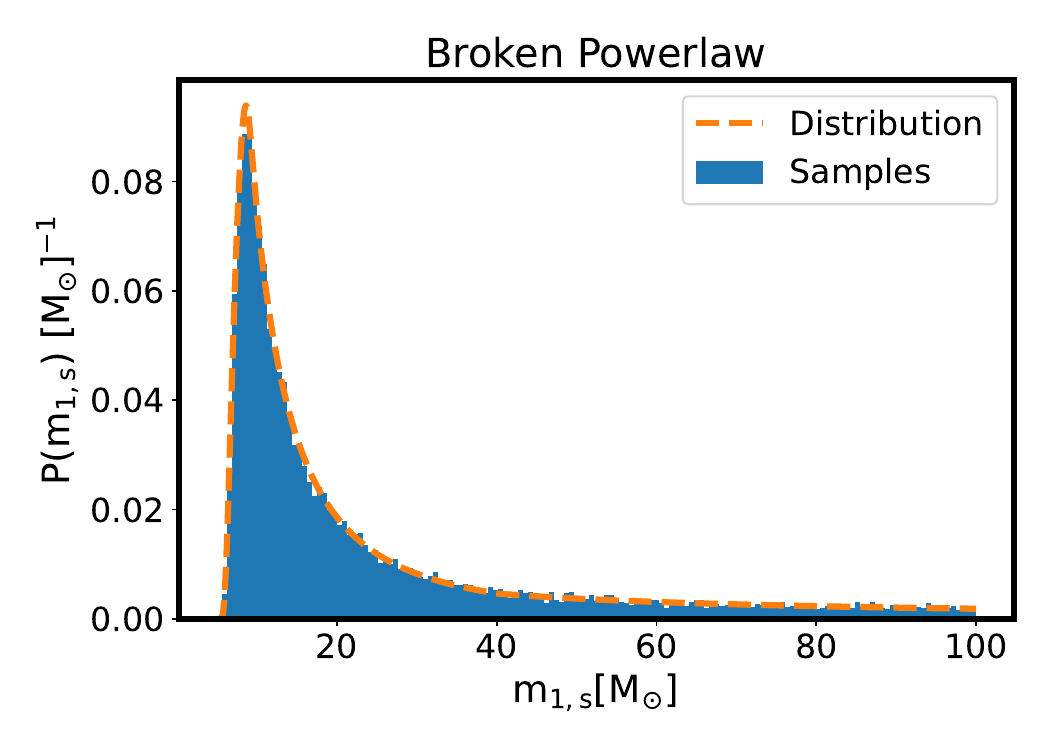}
    \includegraphics[scale=.45]{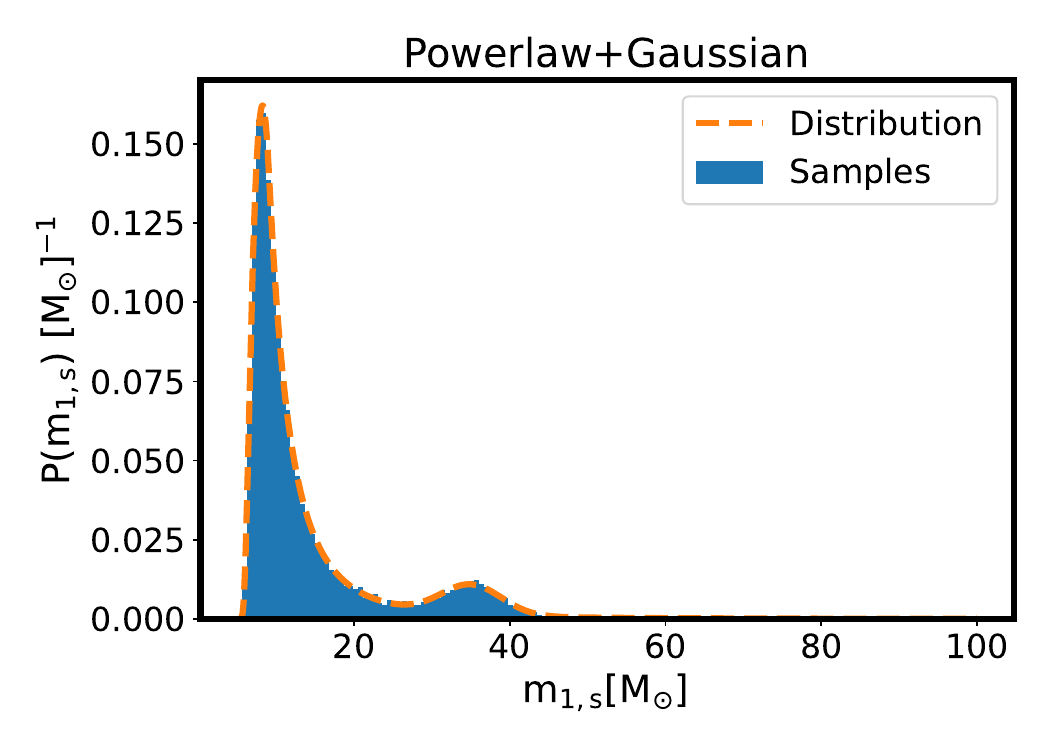}
    \includegraphics[scale=.45]{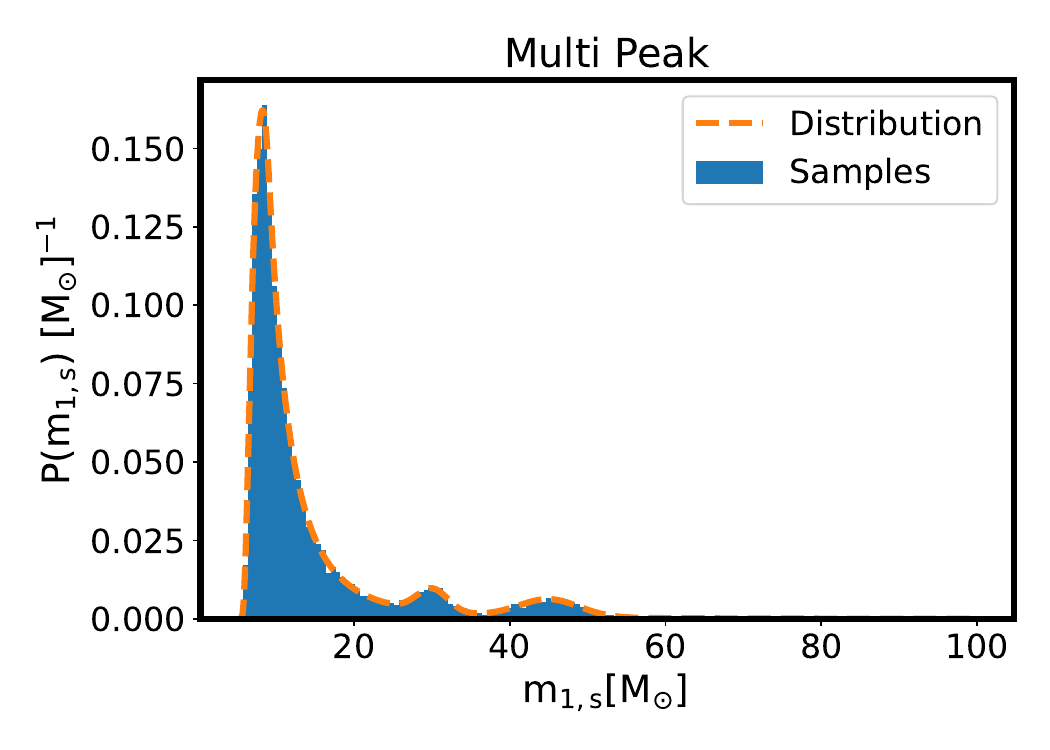}
    \caption{Empirical distributions (in blue) of the source-frame mass samples drawn from various mass distributions. The analytical expressions of the distributions are plotted in orange dashed curves. Top-left: Truncated power-law model. Top-right: broken-power law model with smoothing at the lower end of the distribution. Bottom-left: a power-law+Gaussian peak model with smoothing at the lower end of the distribution. Bottom-right: a multi-peak model with smoothing at the lower end of the distribution.}
    \label{fig:mass_samples}
\end{figure*}

\section{ {Merger rate distributions}}\label{app:2}
Here we show some of the available merger rate distributions and showcase the ability of the code to properly sample those distributions. 
\subsection{Star formation rate-like BBH merger model}
In this case, the distribution is similar  to  that  of~\cite{Madau2014}, which is  characterized by a low-redshift power-law slope, $\gamma$,  a peak at redshift, $z_p$, and a high-redshift power-law slope, $\kappa$, after the peak:
\begin{equation}\label{eq:madau}
    R(z) = \dder{N}{t_s}{V_c} = R_0  \left[1+(1+z_p)^{-\gamma-\kappa}\right]\dfrac{(1 +z)^{\gamma}}{1 + \left[(1 +z)/(1 +z_p)\right]^{\gamma+\kappa}}.
\end{equation}
\subsection{Time delay BBH merger rate}
In the model that takes into account the time delay, the BBH merger rate at a merging redshift $z_m$ is given by:
\begin{equation}\label{eq:R_z}
    R_{TD}(z_m) = R_0\frac{\int_{z_m}^{\infty} P_t(t_d|t_d^{min},t_d^{\rm max},d)  R(z_f)\der{t}{z_f}\dd z_f}{\int_{0}^{\infty} P_t(t_d|t_d^{min},t_d^{\rm max},d)  R(z_f)\der{t}{z_f}\dd z_f},
\end{equation}
where $R_0 \equiv R(z=0)$ is defined as the merger rate at $z=0$.

In Fig.~\ref{fig:redshift_samples}, the samples drawn from the distributions are histogrammed in blue and the analytical curves are shown as orange dashed curves.
\begin{figure*}[!ht]
    \centering
    \includegraphics[scale=.45]{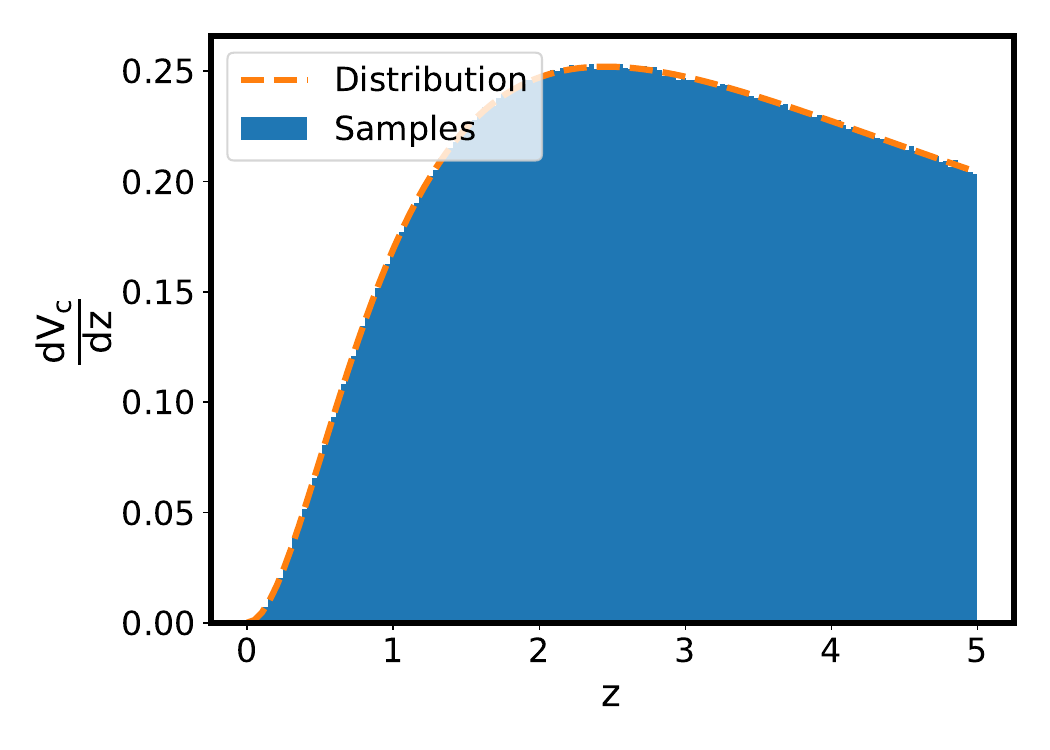}
    \includegraphics[scale=.45]{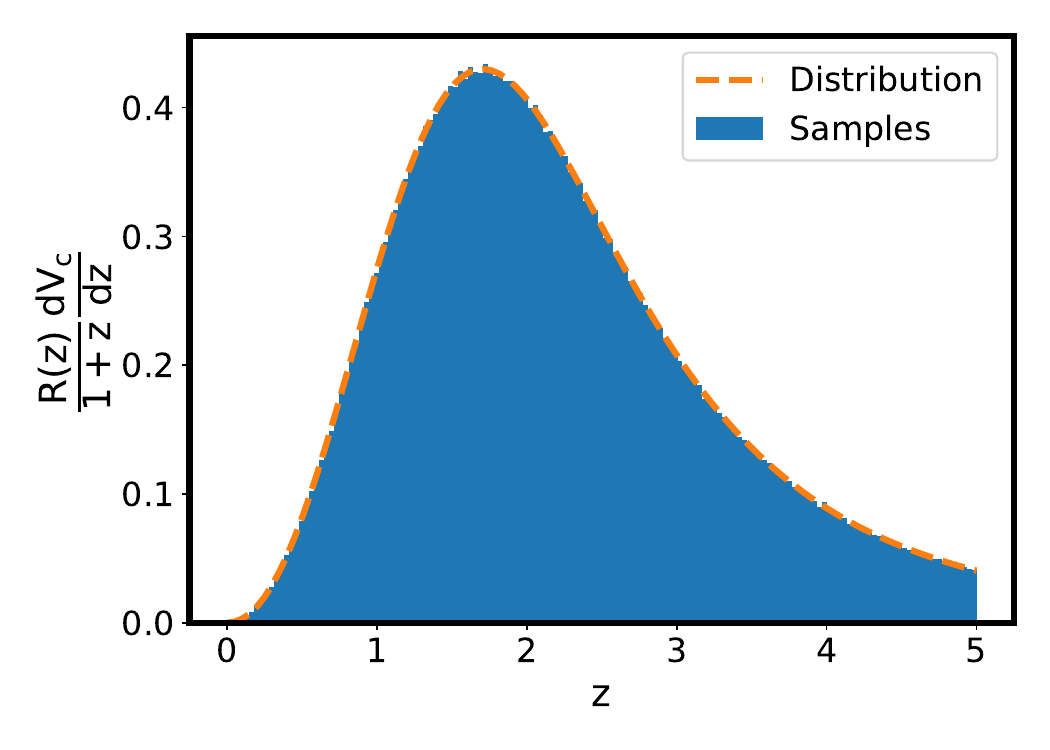}
    \begin{center}
        \includegraphics[scale=.45]{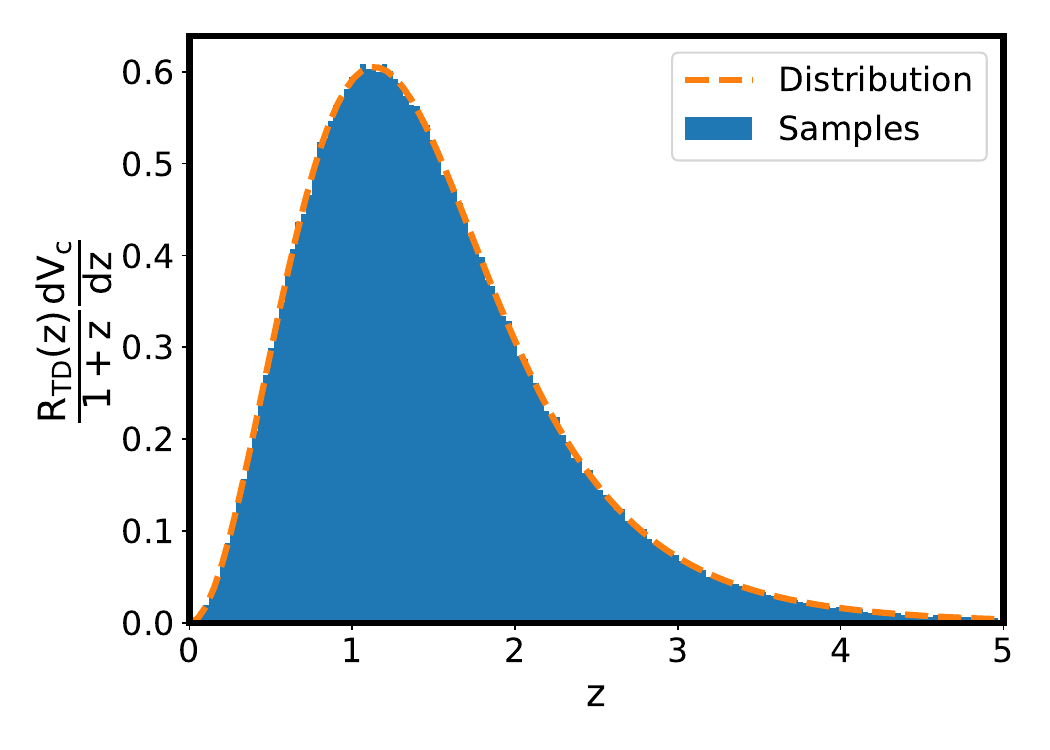}
    \end{center}
    \caption{Empirical distribution of the redshift samples draws (in blue) from the uniform in comoving volume distribution (top left), the merger rate model of \eq~\ref{eq:rz-p} (top right) for $z_p=1.9,\gamma=2.7,\kappa=2.9$, and the merger rate model in the case of time delay model described by \eq~\ref{eq:delay} (bottom) for $d=-1,t_d^{min}=0.5 \ \rm{Gyrs}$. The analytical expressions of the distributions are plotted in orange dashed curves. Results obtained using the parameters values given in \tabacro~\ref{tab:fixedparams}.}
    \label{fig:redshift_samples}
\end{figure*}

\section{ {Spin distributions}}\label{app:3}
Various spin distributions are available within the code for both (i) redshift-independent models and (ii) redshift-dependent models. Here, we discuss the available distributions.
\subsection{Redshift-dependent spin model}\label{app:spin-rid}
\subsubsection{Uniform distribution}
The simplest implemented spin distribution is the uniform one where:
\begin{equation}
    P(\chi_{\rm eff})\propto \mathcal{U}[-1,1],
\end{equation}
and where $\mathcal{U}$ is the uniform distribution.
\subsubsection{Heavy-mass model}
This distribution allows for a correlation of the spins with the masses of the black holes. In this case, individual spins of the black holes are drawn from:
\begin{equation}
    P(\chi|m_{\rm th}) \propto 
        \begin{cases}
            \mathcal{U}[-1,1] & \text{ for $m>m_\text{th}$},\\
            0 & \text{for $m<m_\text{th}$}.\\
        \end{cases}
\end{equation}
This model allows for black holes with masses higher than a selected mass threshold $m_\text{th}$ to have non-zero spins.
\subsubsection{Gaussian spins}
Here $\chi_{\rm eff}$ is taken from a Gaussian distribution:
\begin{equation}
    P(\chi_{\rm eff}|\sigma_{\chi,0},\mu_{\chi,0})\propto G(\chi_{\rm eff}|\sigma_{\chi_{eff}},\mu_{\chi_{eff}}),
\end{equation}
where $G$ is the Gaussian distribution as
\begin{equation}\label{eq:s5}
    G(\chi_{\rm eff}|\sigma_{\chi,0},\mu_{\chi,0})=\dfrac{1}{\sigma_{\chi,0}\sqrt{2\pi}}\exp\left(-\dfrac{(\chi_{\rm eff}-\mu_{\chi,0})^2}{2\sigma_{\chi,0}^2}\right).
\end{equation}

\subsubsection{Mass correlated model}
This model describes a more complicated correlation between masses and spins. In this case, we have:
\begin{align}
\begin{split}
p(\chi_\text{eff}|q,\mu_{\chi,0},\sigma_{\chi,0},&\alpha_s,\beta_s)
 \propto \\ &\dfrac{1}{\sigma_{(\sigma_{\chi,0},\beta_s,q)}\chi\sqrt{2\pi}}\exp
\left[-
\frac{ \left(\chi_\text{eff}-\mu_\chi(\mu_{\chi,0},\alpha_s,q)\right)^2}
{ 2\sigma_{\chi}^2(\sigma_{\chi,0},\beta_s,q)}\right],
\end{split}
\end{align}
where $\mu_\chi=\mu_{\chi,0}+\alpha_s(q-1/2)$ and $\sigma_{\chi}^2(\sigma_{\chi,0},\beta_s,q)$ such that $\log_{10} \sigma_\chi(\sigma_{\chi,0},\beta_s,q)=\log_{10} \sigma_{\chi,0}+\beta_s(q-1/2)$. 
Here we show a spin sampling of the mass correlated model. In Fig.~\ref{fig:spin} one can see the samples that the code drew in the $\xeff-q$ space,  $\chi_{1}-q$ space, and  $\chi_{2}-q$ space. Figure~\ref{fig:spinz} shows the individual spin amplitudes $\chi_1$ and $\chi_2$ as a function of $z$ in the case of the spin-redshift model where both the expectation value and standard deviation depend linearly of~$z$ (see Sect.~\ref{sec:spinz}).
\begin{figure*}[!ht]
    \centering
    \includegraphics[scale=.15]{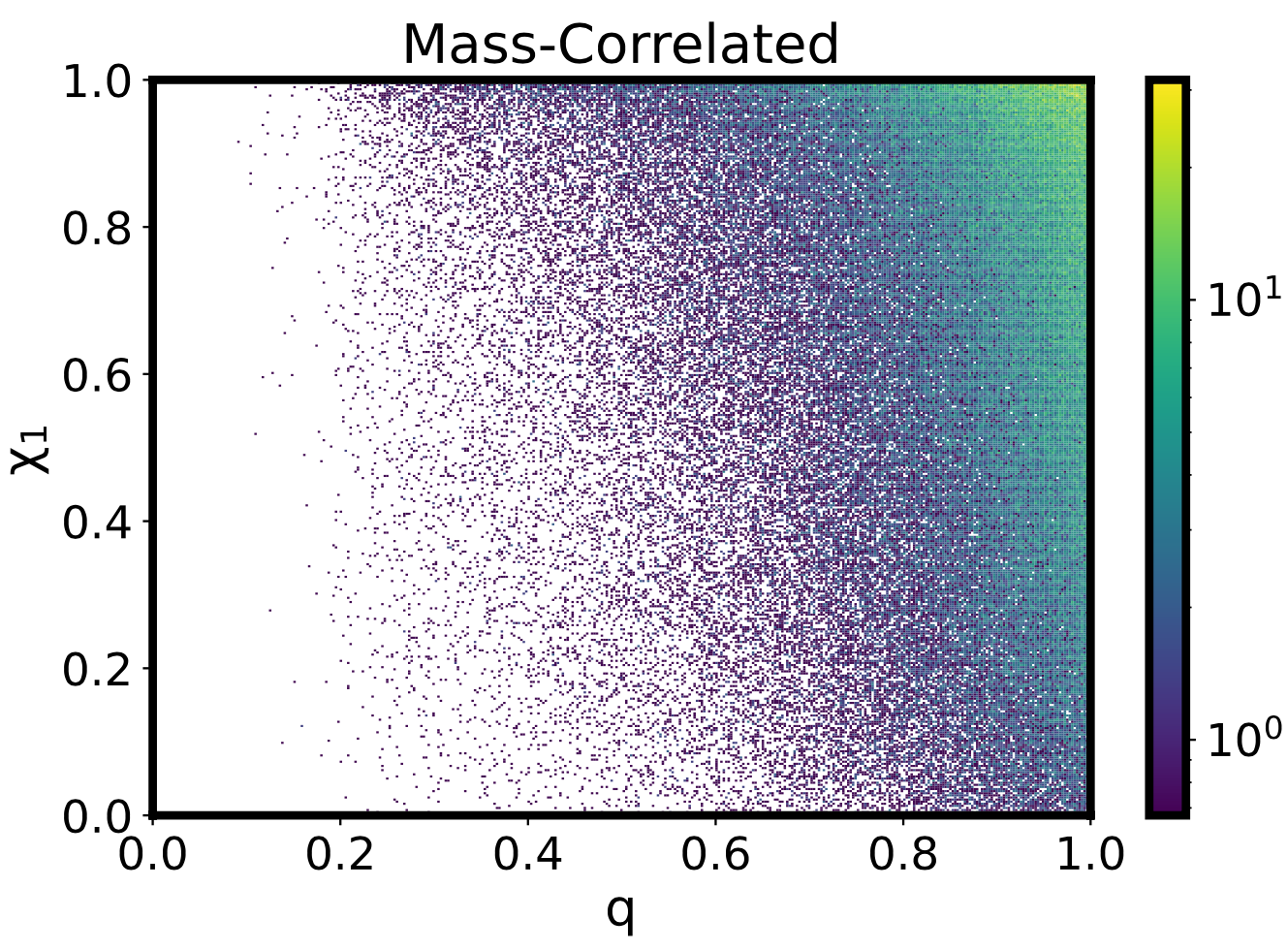}
    \includegraphics[scale=.15]{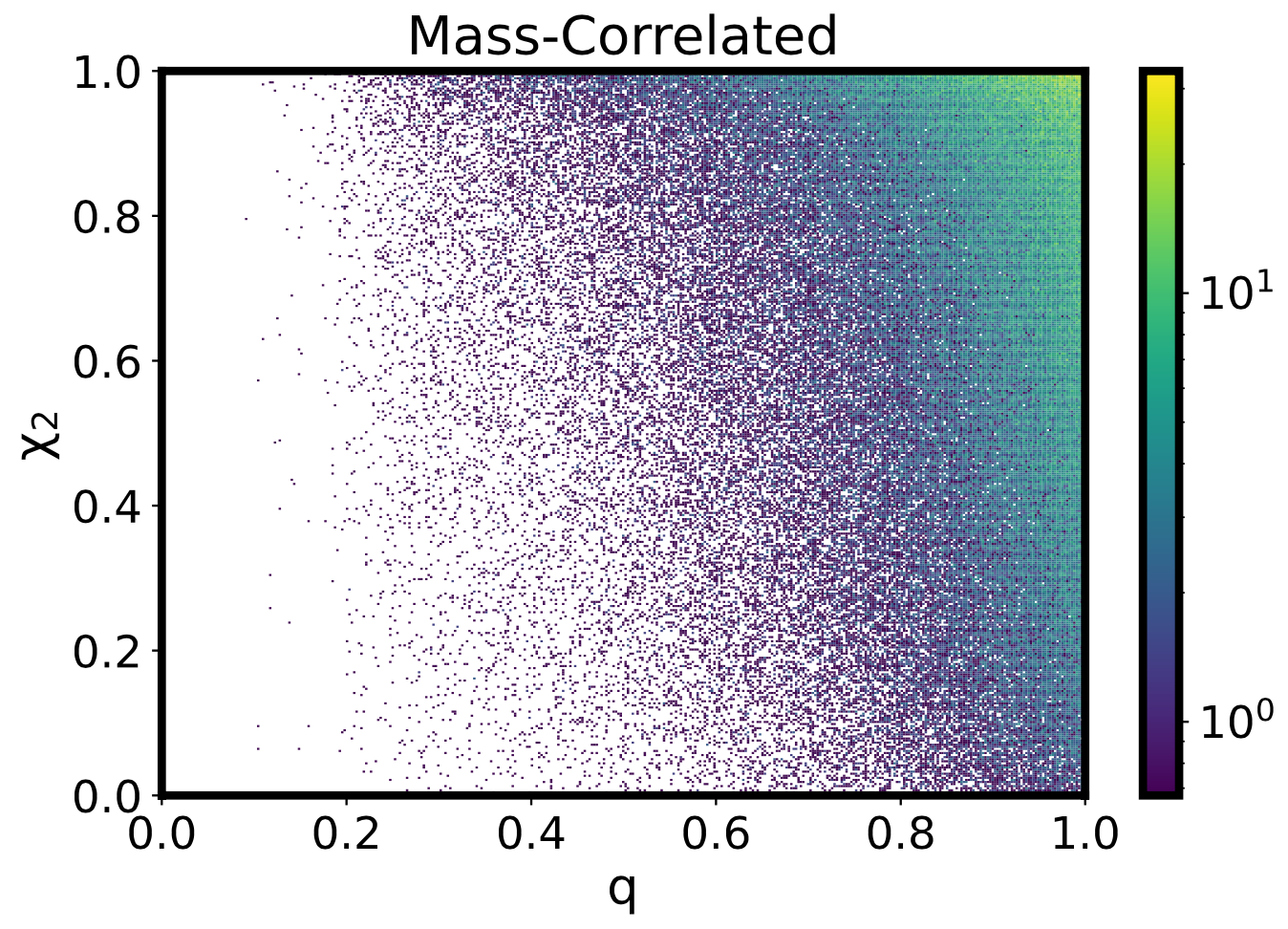}
    \begin{center}
        \includegraphics[scale=.15]{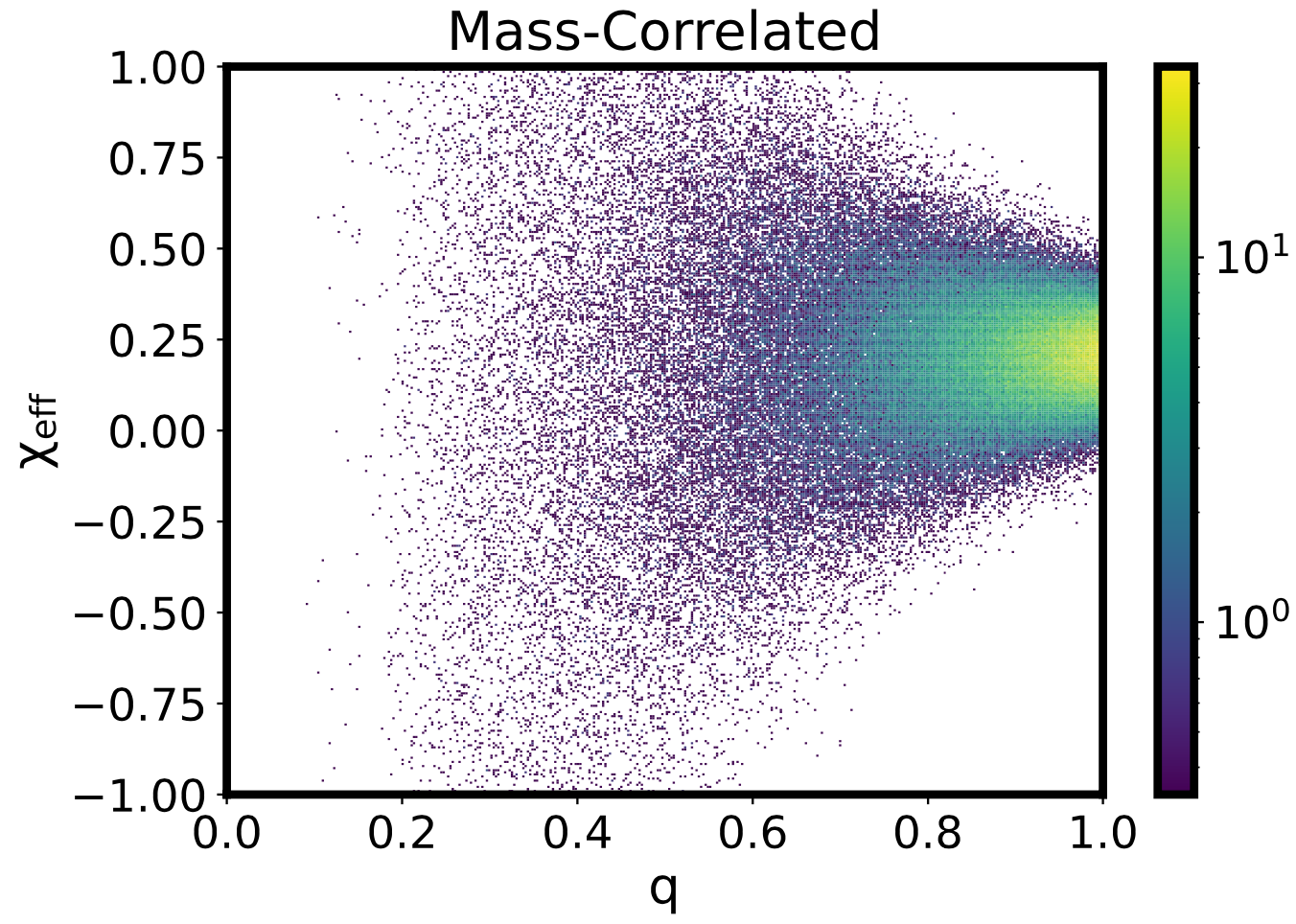}
    \end{center}
    \caption{2D distribution of the spin sampling for the case of the mass correlated model. Top figures: Individual spin amplitudes $\chi_1$ (left) and $\chi_2$ (right) as a function of the mass ratio $q$. Bottom: Effective spin amplitude, $\xeff$, as a function of the mass ratio, $q$. The distributions are calculated with $\alpha_s= 0.013, \beta_s= -1.53, \chi_0= 0.2, \sigma_0= 0.52$. The mass distribution is the power-law+Gaussian distribution seen in \ref{fig:mass_samples}.}
    \label{fig:spin}
\end{figure*}

\begin{figure*}[!ht]
        \begin{center}
        \includegraphics[scale=.5]{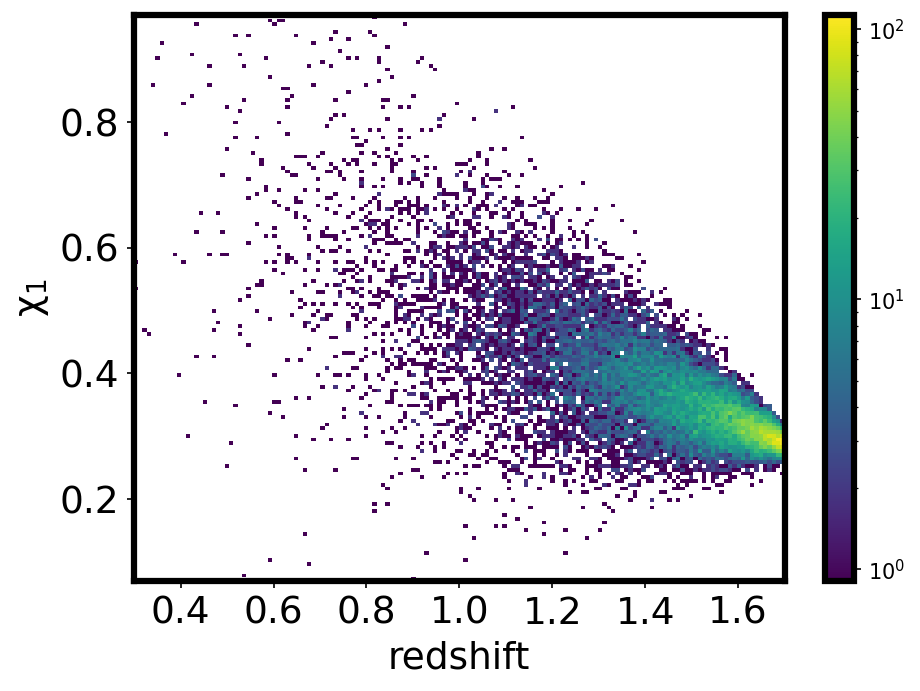}
        \includegraphics[scale=.5]{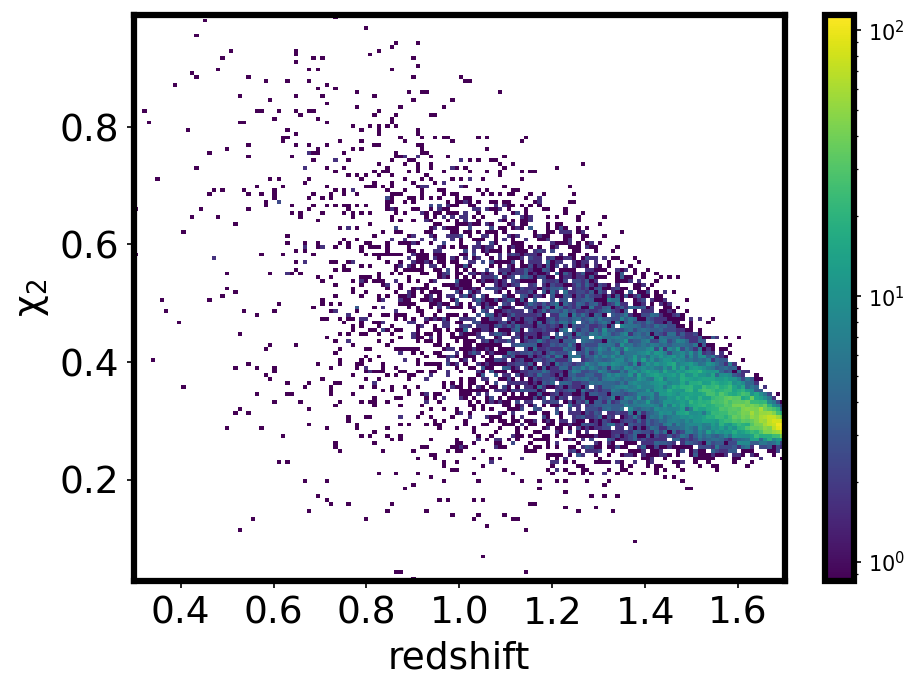}
            \end{center}
    \caption{2D distribution of the sampled spin amplitudes $\chi_1$ (left) and $\chi_2$ (right) for the case of the spin model evolving with the redshift. The parameters used are $\chi_0=0.8$, $\delta\chi=-0.3$, $\sigma_0=0.3$, $\delta\sigma=-0.17$ (see Sect.~\ref{sec:spinz} for the spin-redshift dependence).}
    \label{fig:spinz}
\end{figure*}
\subsection{Redshift-dependent spin model}\label{app:spin-rd}

For the redshift-dependent spin model, we considered the redshift-dependent Gaussian model.  We allowed for a redshift evolution of the mean value of effective spin $\mu_{\chi, 0}(z)$ and standard deviation of $\sigma_{\chi, 0}(z)$ as:
\begin{align}
\begin{split}
    \mu_{\chi, 0}(z)= \mu_{\chi, 0}+ \delta \chi z,\\
    \sigma_{\chi, 0}(z)= \sigma_{\chi, 0}+ \delta \sigma z.
\end{split}
\end{align}
The value of $\chi_{\rm eff}$ is taken from a Gaussian distribution, as follows:
\begin{equation}\label{eq:s5z}
    G(\chi_{\rm eff}|\sigma_{\chi,0}(z),\mu_{\chi,0}(z))=\dfrac{1}{\sigma_{\chi,0}(z)\sqrt{2\pi}}\exp\left(-\dfrac{(\xeff-\mu_{\chi,0}(z))^2}{2\sigma_{\chi,0}(z)^2}\right).
\end{equation}
\label{lastpage}

\end{document}